\documentclass[useAMS,usenatbib,onecolumn]{mn2e}
\usepackage{amssymb}

\title[Ellipsoidal Universe \& Planck]{ The Ellipsoidal Universe in the Planck Satellite Era }

\author[Paolo  Cea]{Paolo  Cea$^{1,2}$\thanks{E-mail:paolo.cea@ba.infn.it)}\\
$^{1}$Dipartimento di Fisica, Universit\`{a} di Bari, I-70126 Bari, Italy\\
$^{2}$INFN - Sezione di Bari, I-70126 Bari, Italy}

\begin{document}

\date{}

\pagerange{\pageref{firstpage}--\pageref{lastpage}} \pubyear{2014}

\maketitle

\label{firstpage}

\begin{abstract}
Recent Planck data confirm that the cosmic microwave background displays the quadru-pole power suppression
together with large scale anomalies. Progressing from previous results, that focused on the quadrupole anomaly, we strengthen the proposal 
that the slightly anisotropic  ellipsoidal universe may account for these anomalies. We solved at large scales the Boltzmann equation for the photon
distribution functions by taking into account both the effects of the inflation produced primordial scalar perturbations
and the anisotropy of the geometry in the ellipsoidal universe. We showed that the low quadrupole temperature correlations
allowed us to fix the eccentricity at decoupling,  $e_{\rm dec} \, = \, ( 0.86 \, \pm \, 0.14) \, 10^{-2}$, and to constraint the direction
of the symmetry axis. We found that the anisotropy of the geometry of the universe contributes only to the large scale temperature anisotropies
without affecting the higher multipoles of the angular power spectrum. Moreover, we showed that the ellipsoidal geometry of the
universe induces sizable polarization signal at large scales without invoking the reionization scenario. We explicitly
evaluated the quadrupole TE and EE correlations. We found an average large scale polarization
 $\Delta T_{pol} \,  =  \, (1.20 \, \pm  \,  0.38) \;  \mu K $. We point out that great care is needed in the
 experimental determination of the large-scale polarization correlations since the average temperature polarization
 could be misinterpreted as foreground emission leading, thereby,  to a considerable underestimate of
 the cosmic microwave background polarization signal.
\end{abstract}


\section{Introduction}
The Cosmic Microwave Background (CMB) anisotropy data produced by the final analysis of the Wilkinson Microwave
Anisotropy Probe (WMAP)~\citep{Bennett:2013,Hinshaw:2013} and, more recently, by the Planck satellite~\citep{Ade:2013a,Ade:2013c,Ade:2013d}  confirme the standard
cosmological Lambda Cold Matter ($\Lambda$CDM) model at an unprecedented level of accuracy. 
At large scales, however, several anomalous features have been reported:  an unusual alignment of the preferred axes of the quadrupole and 
octopole~\citep{Land:2005, Oliveira:2004, Ralston:2004, Copi:2006}, non-Gaussian signatures due to a cold spot~\citep{Cruz:2005},
an  hemispherical power asymmetry  at large scales~\citep{Eriksen:2004, Hansen:2004}. Nevertheless, we feel that  one of the most important discrepancy resides
 in the low quadrupole moment, which signals an important
suppression of power at large scales. In fact, the Planck collaboration reported  a statistical significant  tension between the best fit $\Lambda$CDM model 
and the large-scale spectrum due to a systematic lack of power for $ \ell \, \lesssim \, 40$~\citep{Ade:2013c}  and to anomalies in the statistical isotropy of the
sky maps ~\citep{Ade:2013e}.  \\
If these anomalies should turn out to have a cosmological origin, then it could have far reaching consequences for
our present understanding of the universe. \\
\indent
Quite recently it has been suggested~\citep{Campanelli:2006,Campanelli:2007,Cea:2010} that, 
if one admits that the large-scale spatial geometry of our universe 
is only  plane-symmetric with eccentricity at decoupling of order $10^{-2}$, 
then the quadrupole amplitude can be drastically reduced without affecting higher multipoles of the angular power spectrum of the temperature anisotropy.  
As discussed in ~\citet{Campanelli:2007},  the anisotropic expansion described by a plane-symmetric metric can be generated by cosmological 
magnetic fields or topological defects, such as cosmic domain walls or cosmic strings. 
Indeed, topological cosmic defects are relic structures that are predicted to be produced in the course of symmetry breaking in the hot, early universe (e.g., see \citet{Vilenkin:1994}).  \\
\subsection{Outline of the Results}
In an isotropic and homogeneous universe the most general  metric  is the Friedmann-Robertson-Walker (FRW) metric (see, for instance, \citealt{Peebles:1993}).
In particular, the metric of standard cosmological model is given by ~\footnote{Note that  through the paper we shall use units in which $c \; = \; 1$, $\hbar \; = \; 1$ and $k_B \; = \; 1$.}:
\begin{equation}
\label{1.1}
ds^2 = - dt^2 + a^2(t) \delta_{ij}  \, dx^i dx^j \; .
\end{equation}
If we assume that  the large-scale spatial geometry of our universe is only  plane-symmetric, then the metric  Eq.~(\ref{1.1})  is replaced with the ellipsoidal universe metric: 
\begin{equation}
\label{1.2}
ds^2 = - dt^2 + a^2(t) (\delta_{ij} + h_{ij}) \, dx^i dx^j \; ,
\end{equation}
where $h_{ij}$ is a metric perturbation which we assume to be of the form:
\begin{equation}
\label{1.3}
h_{ij} \; = \;  - \, e^2(t) \;  n_{i}  \, n_{j} \; .
\end{equation}
In Eq.~(\ref{1.3}) $e(t) = \sqrt{1 - (b(t)/a(t))^2} $ is the ellipticity and the unit vector $\vec{n}$ determines the direction of the symmetry axis. \\
In this paper we shall further elaborate on the  ellipsoidal universe proposal and extend previous investigations in several directions.
For reader's convenience, it is useful to summarize the main results of the present paper. \\
First, we consider the Boltzmann equation for the photon distribution in the ellipsoidal universe, discussed for the first time
in \citet{Cea:2010}, by taking into account also the effects of the cosmological  inflation produced primordial scalar perturbations.
In the large scale approximation we explicitly show that the CMB temperature fluctuations can be written as:
\begin{equation}
\label{1.4}
\Delta T  \;  \simeq \; \Delta T^I  \;   +  \;   \Delta T^A  \;  ,
\end{equation}
where   $\Delta T^I$ and   $ \Delta T^A$ are the temperature fluctuations  induced by the cosmological scalar perturbations and by the spatial anisotropy
of the metric of the universe, respectively.  Since the temperature anisotropies caused by the  inflation produced primordial scalar perturbations
are discussed in several textbooks~\citep{Dodelson:2003,Mukhanov:2005}, we focus on the temperature fluctuations  induced by the
anisotropy of the metric by solving the relevant   Boltzmann equation.
At large scales we solve that equation and determine the solutions relevant for the CMB temperature and polarization fluctuations. Indeed, it
is well known~\citep{Rees:1968, Negroponte:1980,Basko:1980} that anisotropic cosmological models 
give sizable contributions to the large scale polarization of the cosmic microwave background radiation. 
In fact, polarization measurements could provide a unique signature of cosmological anisotropies. \\
We go beyond the approximations adopted in \citet{Cea:2010} and confirm that the main contributions to 
the CMB temperature fluctuations affect  the quadrupole correlations. In addition, we also show that the
effects of the spatial anisotropy of the metric of the universe extend to low-lying multipoles $\ell \sim 10$. \\
As is well known, the CMB temperature fluctuations are fully characterized by the power spectrum:
\begin{equation}
\label{1.5}
( \Delta T_{\ell} )^2  \;  \equiv  \; \mathcal{D}_{\ell}   \; = \; \frac{\ell(\ell+1)}{2 \pi}  \; C_{\ell} \; \; , \; \; 
C_{\ell} \; = \; \frac{1}{2 \ell+1} \sum_{m=-\ell}^{+\ell} \,  | a_{\ell m} |^2 \; .
\end{equation}
In particular, the quadrupole anisotropy refers to the multipole $\ell = 2$. Remarkably, the Planck data~\citep{Ade:2013c} 
confirmed that the observed quadrupole anisotropy:
\begin{equation}
\label{1.6}
( \Delta T_{2} )^2  \;  =  \; \mathcal{D}_{2}   \; \simeq \; 
299.5 \;  \;  \mu \, K^2 \;  , 
\end{equation}
is much smaller than the quadrupole anisotropy expected  according to the best fit  $\Lambda$CDM model to the Planck data:
\begin{equation}
\label{1.7}
( \Delta T^I_{2} )^2  \;  =  \;  1150 \; \pm \; 727 \;   \mu \, K^2 \; .
\end{equation}
Note that in Eq.~(\ref{1.6}) we are neglecting the rather small measurement errors, while the uncertainties due
to the so-called cosmic variance are included in the theoretical expectations, Eq.~(\ref{1.7}).  
In fact, using Eq.~(\ref{1.4}) we show that the quadrupole temperature anisotropy can be reconciled  with observations
in the ellipsoidal universe if the eccentricity at decoupling is:
\begin{equation}
\label{1.8}
e_{\rm dec} \;  =  \; ( 0.86 \;  \pm \; 0.14) \; 10^{-2} \; ,
\end{equation}
irrespective of the physical mechanism responsible for the generation of the spatial anisotropy in the early universe. Moreover,
if we denote with $b_n$ and $l_n$  the galactic latitude and longitude of the symmetry axis respectively, we  also
were able to show that   the axis of symmetry were   constrained to:
\begin{equation}
\label{1.9}
b_n  \; \simeq \;  \pm 17^{\circ}  \;  \; ,
\end{equation}
while the  longitude $b_n$ turns out to be  poorly constrained, in qualitative agreement with  \citet{Campanelli:2007}. \\
As concern the CMB polarization, we confirm our previous result ~\citep{Cea:2010} that  the ellipsoidal geometry of the
universe induces sizable polarization signal at large scales without invoking the reionization scenario.
In particular, we find an average large scale polarization:
\begin{equation}
\label{1.10}
 \Delta T_{pol} \; \equiv \; \frac{1}{4 \pi} \; \int \; d\Omega \; \;  \Delta T^{E}(\theta,\phi)  \;  
 =  \; (1.20  \, \pm  \,  0.38) \;  \mu K \; , 
\end{equation}
where  $\Delta T^{E}(\theta,\phi)$ is the polarization of the CMB  temperature fluctuations.
Moreover, we evaluate the quadrupole temperature-polarization cross-correlation (TE) and   polarization-polarization (EE) correlation.  We find:
\begin{equation}
\label{1.11}
\Delta T^{TE}_{2}   \;  =  \;  3.14 \; \pm \; 0.76  \;   \mu \, K \; ,
\end{equation}
and
\begin{equation}
\label{1.12}
\Delta T^{EE}_{2}   \;  =  \;  0.83 \; \pm \; 0.27  \;   \mu \, K \; .
\end{equation}
These values should be compared with the available observational data.  Since the Planck collaboration does not yet
make public the large scale polarization data, we must rely on the final analysis of the Wilkinson Microwave
Anisotropy Probe collaboration. The WMAP nine-year full-sky maps 
of the polarization detected at large scales in the foreground corrected maps an average E-mode polarization power~\citep{Bennett:2013,Hinshaw:2013}.
In particular for the quadrupole correlations we have (including only the statistical uncertainties):
\begin{equation}
\label{1.13}
\frac{l(l+1)}{2 \pi} \; C^{TE}_{l= 2} \; = \;  2.4439 \; \pm  2.2831    \; \;  \mu \, K^2  \;  \; \; \; , \;  \; WMAP  \; \; nine-years
\end{equation}
and
\begin{equation}
\label{1.14}
\frac{l(l+1)}{2 \pi} \; C^{EE}_{l= 2} \; = \; -0.0860  \; \pm  0.0247    \; \;  \mu \, K^2  \; , \;  \; WMAP  \;  \; nine-years
\end{equation}
Using the definition in Eq.~(\ref{1.5}) we estimate:
\begin{equation}
\label{1.15}
\Delta T^{TE}_{2}   \;  =  \;  1.56 \; \pm \; 0.73 \;   \mu \, K  \;  \; ,  \;  \; WMAP \; nine-years
\end{equation}
which within two standard deviations agrees with our result Eq.~(\ref{1.11}). On the other hand, as concern the quadrupole EE correlation,
Eq.~(\ref{1.14}) at best gives an upper bound which, however, is not consistent with our result  Eq.~(\ref{1.12}). 
We believe that this discrepancy could be due to the fact that in the ellipsoidal universe model,  
at variance of the standard reionization scenario, there is a non-zero average temperature polarization.
In fact, the eventual presence of an average temperature polarization could be misinterpreted 
as foreground emission leading  to an underestimate of the cosmic microwave background polarization signal. \\
\indent
The plan of the paper is as follows. In sect.~\ref{S2} we discuss the Boltzmann equation of the cosmic background radiation in the ellipsoidal
universe. In sect.~\ref{S3} we determine the solutions of the Boltzmann equation at large scales. Sect.~\ref{S4} is devoted to the problem of the quadrupole anomaly
in the temperature-temperature fluctuation correlations. In sect.~\ref{S5} we discuss the large scale polarization. In particular, we determine the quadrupole TE and EE
 correlations.  Finally, our conclusions are drawn in  sect.~\ref{S6}. Some technical details are relegated in  appendix  \ref{Appendix A}, while in appendix
 \ref{Appendix B}  we discuss the multipole expansion of the large scale temperature anisotropies.
\section[]{The Boltzmann equation in the ellipsoidal universe}
\label{S2}
We are interested in the temperature fluctuations of the cosmic background radiation induced by eccentricity of the universe and by the inflation produced primordial cosmological
perturbations. We assume that  the photon distribution function $f(\vec{x},t)$ is an isotropically radiating blackbody at a sufficiently early epoch. 
The subsequent evolution of  $f(\vec{x},t)$ is determined by the Boltzmann equation~\citep{Dodelson:2003,Mukhanov:2005}:
\begin{equation}
\label{2.1}
\frac{df}{dt} \; = \; \left ( \frac{\partial f} {\partial t} \right )_{coll} \; , 
\end{equation}
where $ (\frac{\partial f} {\partial t} )_{coll} $ is the collision integral which takes care of Thomson scatterings between matter and radiation.  \\
The metric of the standard FRW universe is:
\begin{equation}
\label{2.2}
ds^2 = - dt^2 \; + \; a^2(t) \;  \delta_{ij}  \, dx^i dx^j \; .
\end{equation}
Here, we are interested in primordial scalar perturbations induced by the inflation. In the conformal Newtonian, or longitudinal gauge~\citep{Mukhanov:2005}, 
the metric Eq.~(\ref{2.2}) can be written as:
\begin{equation}
\label{2.3}
ds^2 = -  [1 \, + \, 2 \Psi (\vec{x},t)] \, dt^2 \; + \; a^2(t) \;  \delta_{ij}  \, [1 \, + \, 2 \Phi (\vec{x},t)]  \, dx^i dx^j \; .
\end{equation}
In this gauge the perturbations to the metric are determined by the functions $\Psi (\vec{x},t)$ and  $\Phi (\vec{x},t)$ which correspond
to the Newtonian potential and the perturbation to the spatial curvature, respectively. In the ellipsoidal universe the metric would be:
\begin{equation}
\label{2.4}
ds^2  =  - [1 \, + \, 2 \Psi (\vec{x},t)] \, dt^2  \;  + \; a^2(t) [\delta_{ij} + h_{ij}] \,  [1 \, + \, 2 \Phi (\vec{x},t)] \, dx^i dx^j \; ,
\end{equation}
where $h_{ij}$ is given by Eq.~(\ref{1.3}). However, both the primordial perturbations  $\Psi (\vec{x},t)$,  $\Phi (\vec{x},t)$ and
the ellipticity are to be considered small at the times and scales of interest. Therefore in the following we shall neglect all terms quadratic in them.
Accordingly, instead of Eq.~(\ref{2.4}) we have:
\begin{equation}
\label{2.5}
ds^2  =  - [1 \, + \, 2 \Psi (\vec{x},t)] \, dt^2  \;  + \; a^2(t)  \; \{  \delta_{ij} \,  [1 \, + \, 2 \Phi (\vec{x},t)] \, + \,  h_{ij} \} \,   dx^i dx^j \; .
\end{equation}
We are interested in the anisotropies in the cosmic distribution of photons. To this end, we need to evaluate the photon distribution function
$f(\vec{x},t)$ which satisfies the Boltzmann equation Eq.~(\ref{2.1}). Actually, the distribution function depends on the space-time point $x^{\mu}$ and
the momentum vector   $p^{\mu}$ defined by: 
\begin{equation}
\label{2.6}
 p^{\mu} \;  =  \;  \frac{d \, x^{\mu}}{d \, \lambda} \; ,
\end{equation}
where $\lambda$ parametrizes the particle's path. For massless particles we, obviously, have:
\begin{equation}
\label{2.7}
 P^2 \;  \equiv   \;  g_{\mu \, \nu} \;    p^{\mu} \, p^{\nu} \;  = \; 0 \; .
\end{equation}
Using the metric in   Eq.~(\ref{2.5}) and defining:
\begin{equation}
\label{2.8}
 p^2 \;  \equiv   \;  g_{i \, j} \;    p^{i} \, p^{j}  \;  ,
\end{equation}
from  Eq.~(\ref{2.7}) we easily obtain:
\begin{equation}
\label{2.9}
 p^0 \; \simeq  \;  p \; [ 1 \, - \, \Psi ]    \;  .
\end{equation}
It is convenient to consider the distribution function as a function of the magnitude of momentum $p$ and 
momentum direction $\hat{p}^i$, $ \delta_{ij} \,  \hat{p}^i \hat{p}^j = 1$. Therefore we have:
\begin{equation}
\label{2.10}
\frac{df}{dt} \; = \; \frac{\partial f}{\partial t} + \frac{\partial f}{\partial x^i} \,   \frac{d x^i}{d t} +  \frac{\partial f}{\partial p} \,   \frac{d p}{d t} 
+  \frac{\partial f}{\partial \hat{p}^i} \,   \frac{d \hat{p}^i}{d t}  \; .
\end{equation}
Now we note that:
\begin{equation}
\label{2.11}
  \frac{d x^i}{d t}  \; = \;  \frac{d x^i}{d \lambda}  \,  \frac{d \lambda}{d t}   \; = \;   \frac{p^i}{p^0}    \; .
\end{equation}
Let us write
\begin{equation}
\label{2.12}
 p^i  \; = \; C \,  \hat{p}^i    \; ,
\end{equation}
then it is easy to find:
\begin{equation}
\label{2.13}
 C  \;  \simeq  \;  \frac{p}{a(t)} \, [ 1 - \Phi - \frac{1}{2} h_{i  j}   p^{i} \, p^{j}  ] \; . 
\end{equation}
So that we have:
\begin{equation}
\label{2.14}
 \frac{d x^i}{d t}  \; \simeq  \; \frac{\hat{p}^i}{a(t)} \, [ 1 - \Phi + \Psi - \frac{1}{2} h_{i  j}   p^{i} \, p^{j}  ] \; . 
\end{equation}
Thus, we get:
\begin{equation}
\label{2.15}
\frac{df}{dt} \; \simeq  \; \frac{\partial f}{\partial t} + \frac{\partial f}{\partial x^i} \,  \frac{\hat{p}^i}{a(t)} \, [ 1 - \Phi + \Psi - \frac{1}{2} h_{i  j}   p^{i} \, p^{j}  ] 
+  \frac{\partial f}{\partial p} \,   \frac{d p}{d t}  +  \frac{\partial f}{\partial \hat{p}^i} \,   \frac{d \hat{p}^i}{d t}  \; \simeq \;
\; \frac{\partial f}{\partial t} + \frac{\partial f}{\partial x^i} \,    \frac{\hat{p}^i}{a(t)}  +  \frac{\partial f}{\partial p} \,   \frac{d p}{d t} 
\end{equation}
since $ \frac{\partial f}{\partial x^i}$ is already a first-order term. To evaluate  $\frac{d p}{d t}$, we note that the time component of the
geodesic equations gives: 
\begin{equation}
\label{2.16}
 \frac{d p^0}{d \lambda}  \; = \; - \; \Gamma^0_{\alpha \beta} \; p^{\alpha} p^{\beta} \; . 
\end{equation}
Since
\begin{equation}
\label{2.17}
 \frac{d p^0}{d \lambda}  \; = \;   \frac{d p^0}{d t}  \,   \frac{d t}{d \lambda}  \; = \;  p^0 \, \frac{d p^0}{d t}   \;  , 
\end{equation}
after using  Eq.~(\ref{2.9}) we obtain:
\begin{equation}
\label{2.18}
 \frac{d p}{d t}  \; \simeq \;  p \, \frac{d \Psi}{d t} \; - \; \Gamma^0_{\alpha \beta} \; \frac{ p^{\alpha} p^{\beta}}{p} \, [1 + 2 \Psi] 
   \; =  \;  p \, [ \frac{\partial \Psi}{\partial t} +  \frac{\hat{p}^i}{a(t)}  \, \frac{\partial \Psi}{\partial x^i} \,   ]  \; - \; \Gamma^0_{\alpha \beta} \; \frac{ p^{\alpha} p^{\beta}}{p} \, [1 + 2 \Psi]   \;  , 
\end{equation}
Moreover, a standard calculation~\citep{Dodelson:2003} shows that:
\begin{equation}
\label{2.19}
\Gamma^0_{\alpha \beta} \; \frac{ p^{\alpha} p^{\beta}}{p}  \; \simeq \;  p \, (1 - 2 \Psi) \,  [  \frac{\partial \Psi}{\partial t} + 2  \frac{\hat{p}^i}{a(t)}  \, \frac{\partial \Psi}{\partial x^i} \, 
 +     \frac{\partial \Phi}{\partial t} +   \frac{1}{2}   \hat{p}^{i} \, \hat{p}^{j}    \frac{\partial h_{i  j} }{\partial t}  + H ]    \;  , 
\end{equation}
where  $H = \dot{a}/a$ is the Hubble rate. Finally, inserting Eqs.~(\ref{2.18}) and  (\ref{2.19}) into  Eq.~(\ref{2.15}) and collecting terms we obtain:
\begin{equation}
\label{2.20}
\frac{df}{dt} \; \simeq \; \frac{\partial f}{\partial t}  \, +  \, \frac{\hat{p}^i}{a(t)} \, \frac{\partial f}{\partial x^i} \, - p \,  \frac{\partial f}{\partial p} \,  
 [ H(t) \, + \,    \frac{\partial \Phi}{\partial t}  \, + \,  \frac{\hat{p}^i}{a(t)}  \, \frac{\partial \Psi}{\partial x^i}  \, + \, 
  \frac{1}{2}   \hat{p}^{i} \, \hat{p}^{j}    \frac{\partial h_{i  j} }{\partial t}  ] \; . 
\end{equation}
To go further we expand the photon distribution about its zero-order Bose-Einstein value:
\begin{equation}
\label{2.21}
f_0(p,t) \; =  \;  \frac{1}{e^{\frac{p}{T(t)}} \, - \, 1} \; .
\end{equation}
We write:
\begin{equation}
\label{2.22}
f(\vec{x},t,p,\hat{p}) =  \frac{1}{e^{\frac{p}{T(t)[1\, + \, \Theta(\vec{x},t,p,\hat{p})]}} \, - \, 1} \; ,
\end{equation}
and expand to the first order in the perturbation  $\Theta(\vec{x},t,p,\hat{p})$:
\begin{equation}
\label{2.23}
f(\vec{x},t,p,\hat{p}) \; \simeq \; f_0(p,t)  \;  - \;  p \,  \frac{\partial f_0}{\partial p}  \, \Theta(\vec{x},t,p,\hat{p})  \; .
\end{equation}
Using the relation $\frac{\partial\,  \ln \, f_0}{\partial \, \ln \, p} \simeq -1$ which is valid in the Rayleigh-Jeans region, we can rewrite  
Eq.~(\ref{2.23}) as:
\begin{equation}
\label{2.24}
f(\vec{x},t,p,\hat{p}) \; \simeq \; f_0(p,t)  \; [ 1 \; + \;  \Theta(\vec{x},t,p,\hat{p}) ]  \; .
\end{equation}
If we neglect the perturbations, it is easy to see that the zero-order Boltzmann equation is satisfied by the Planck distribution
Eq.~(\ref{2.21}) with $T(t) \sim \frac{1}{a(t)}$.  To determine the perturbed distribution $\Theta(\vec{x},t,p,\hat{p})$ we need
to evaluate the Boltzmann equation to the first order.  From Eqs.~(\ref{2.20}) and  (\ref{2.22}) it follows that:
\begin{equation}
\label{2.25}
\left (  \frac{df}{dt} \right )_{first \; order}  \; \simeq \;  - \;  p \,  \frac{\partial f_0}{\partial p}  \, 
\left \{   \frac{\partial \Theta }{\partial t}  \, +  \, \frac{\hat{p}^i}{a(t)} \, \frac{\partial \Theta}{\partial x^i} \, 
 + \,    \frac{\partial \Phi}{\partial t}  \, + \,  \frac{\hat{p}^i}{a(t)}  \, \frac{\partial \Psi}{\partial x^i}  \, + \, 
  \frac{1}{2}   \hat{p}^{i} \, \hat{p}^{j}    \frac{\partial h_{i  j} }{\partial t}   \right \} \; . 
\end{equation}
Thus the first-order Boltzmann equation becomes:
\begin{equation}
\label{2.26}
\frac{\partial \Theta }{\partial t}  \, +  \, \frac{\hat{p}^i}{a(t)} \, \frac{\partial \Theta}{\partial x^i} \, 
 + \,    \frac{\partial \Phi}{\partial t}  \, + \,  \frac{\hat{p}^i}{a(t)}  \, \frac{\partial \Psi}{\partial x^i}  \, + \, 
  \frac{1}{2}   \hat{p}^{i} \, \hat{p}^{j}    \frac{\partial h_{i  j} }{\partial t}   \; \simeq  \;  
 \frac{1}{f_0}  \;     \left ( \frac{\partial f} {\partial t} \right )_{coll}  \; . 
\end{equation}
The collision integral is in general a non linear functional of the distribution function. However, in the first order approximation
it is a linear functional of $\Theta(\vec{x},t,p,\hat{p})$. Moreover, since we are interested in the solutions of 
the Boltzmann equation at large scales, we may neglect the effects due to the bulk velocity of the electrons
which participate to the photon Compton scatterings. In this case the collision integral can be considered a linear
homogeneous functional of the distribution function  $\Theta(\vec{x},t,p,\hat{p})$. As a consequence, if we write
\begin{equation}
\label{2.27}
\Theta(\vec{x},t,p,\hat{p})    \;  \simeq \; \Theta^A(\vec{x},t,p,\hat{p}) \; + \;  \Theta^I(\vec{x},t,p,\hat{p})  \; ,
\end{equation}
then we also have:
\begin{equation}
\label{2.28}
 \left ( \frac{\partial f} {\partial t} \right )_{coll}[\Theta] \; \simeq \;
 \left ( \frac{\partial f} {\partial t} \right )_{coll}[\Theta^A] \; + \;   \left ( \frac{\partial f} {\partial t} \right )_{coll}[\Theta^I]  \;.
\end{equation}
In fact, Eq.~(\ref{2.26}) suggests that we may associate $\Theta^A$ and $\Theta^I$ with the temperature fluctuations 
induced by the spatial anisotropy of the geometry of the universe and by the scalar perturbations
generated during the inflation, respectively.  Accordingly we set:
\begin{equation}
\label{2.29}
\frac{\partial \Theta^I }{\partial t}  \, +  \, \frac{\hat{p}^i}{a(t)} \, \frac{\partial \Theta^I}{\partial x^i} \, 
 + \,    \frac{\partial \Phi}{\partial t}  \, + \,  \frac{\hat{p}^i}{a(t)}  \, \frac{\partial \Psi}{\partial x^i}    \; \simeq  \;  
 \frac{1}{f_0}  \;     \left ( \frac{\partial f} {\partial t} \right )_{coll}[\Theta^I]  \; , 
\end{equation}
and
\begin{equation}
\label{2.30}
\frac{\partial \Theta^A }{\partial t}  \, +  \, \frac{\hat{p}^i}{a(t)} \, \frac{\partial \Theta^A}{\partial x^i} \, 
 + \,   \frac{1}{2}   \hat{p}^{i} \, \hat{p}^{j}    \frac{\partial h_{i  j} }{\partial t}      \; \simeq  \;  
 \frac{1}{f_0}  \;     \left ( \frac{\partial f} {\partial t} \right )_{coll}[\Theta^A]  \; .
\end{equation}
We note that from Eq.~(\ref{2.22}) it follows that the distribution function   $\Theta(\vec{x},t,p,\hat{p})$ is identified 
with the temperature contrast function:
\begin{equation}
\label{2.31}
\Theta(\vec{x},t,p,\hat{p}) \; = \; \frac{\Delta T(\vec{x},t,p,\hat{p})}{T(t)} \;. 
\end{equation}
Therefore, at large scales   Eq.~(\ref{2.27}) implies that:
\begin{equation}
\label{2.32}
\Delta T(\vec{x},t,p,\hat{p})  \; \simeq \;   \Delta T^A(\vec{x},t,p,\hat{p}) \; + \;  \Delta T^I(\vec{x},t,p,\hat{p}) \; . 
\end{equation}
Note that Eq.~(\ref{2.32}) was assumed in ~\citet{Campanelli:2006,Campanelli:2007,Cea:2010}.
Even though this hypothesis was considered reasonable, an explicit proof was lacking. Our discussion shows that Eq.~(\ref{2.32}) 
arises as  a natural consequence of the Boltzmann equation which, however, is valid only at large distances. \\  
We may conclude that to determine the CMB temperature fluctuations at large scales we need to solve the Boltzmann
equations  Eqs.~(\ref{2.29}) and  (\ref{2.30}). Eq.~(\ref{2.29}) is the Boltzmann equation of the standard $\Lambda$CDM cosmological
model, and it has been extensively discussed in several textbooks~\citep{Dodelson:2003,Mukhanov:2005}. Therefore,
in the following we focus on the Boltzmann equation  Eq.~(\ref{2.30}), derived for the first time in 
\citet{Cea:2010}, which allows us to find the CMB temperature fluctuations caused by the anisotropy of the geometry
of the universe.
\section[]{Large scale solutions of the Boltzmann equation}
\label{S3}
In this section we discuss the Boltzmann equation in the ellipsoidal universe Eq.~(\ref{2.30}):
\begin{equation}
\label{3.1}
\frac{\partial \Theta(\vec{x},t,p,\hat{p})   }{\partial t}  \, +  \, \frac{\hat{p}^i}{a(t)} \, \frac{\partial \Theta(\vec{x},t,p,\hat{p})}{\partial x^i} \, 
 + \,   \frac{1}{2}   \hat{p}^{i} \, \hat{p}^{j}    \frac{\partial h_{i  j} }{\partial t}      \; \simeq  \;  
 \frac{1}{f_0}  \;     \left ( \frac{\partial f} {\partial t} \right )_{coll}  \; .
\end{equation}
where, for simplicity,  the superscript A has been dropped. In \citet{Cea:2010} we have discussed the  solutions of Eq.~(\ref{3.1}) by neglecting 
the spatial dependence of the temperature contrast function $\Theta(\vec{x},t,p,\hat{p})$. Here we try to solve   Eq.~(\ref{3.1}) in general.
To do this, we introduce the Fourier transform of the  temperature contrast function:
\begin{equation}
\label{3.2}
 \Theta(\vec{x},t,p,\hat{p})  \; = \;   \int \, \frac{d^3k}{(2 \pi)^3} \; e^{i \, \vec{k} \, \cdot \,  \vec{x} }  \;  \Theta(\vec{k},t,p,\hat{p})  \; .
\end{equation}
Taking into account that the collision integral depends linearly on  $\Theta$, we easily obtain:
\begin{equation}
\label{3.3}
\frac{\partial \Theta(\vec{k},t,p,\hat{p})   }{\partial t}  \, +  \, \frac{i \, \vec{k} \cdot \hat{p}}{a(t)} \, \Theta(\vec{k},t,p,\hat{p}) \, 
 + \,   \frac{1}{2}   \hat{p}^{i} \, \hat{p}^{j}    \frac{\partial h_{i  j} }{\partial t}      \; \simeq  \;  
 \frac{1}{f_0}  \;     \left ( \frac{\partial f} {\partial t} \right )_{coll} [\Theta(\vec{k},t,p,\hat{p})]  \; .
\end{equation}
To determine the polarization of the cosmic microwave background we need the polarized distribution function which, in general, is represented by a column vector whose components are the four Stokes parameters~\citep{Chandrasekhar:1960}. In fact, due to the axial symmetry of the metric only two Stokes parameters need to be considered, namely the two intensities of radiation with electric vectors  in the plane containing $\vec{p}$ and $ \vec{n}$  and perpendicular to this plane respectively. As a consequence,  instead of  Eq.~(\ref{2.24})  we have:
\begin{equation}
\label{3.4}
f(\vec{x},t,p,\hat{p}) \; \simeq \; f_0(p,t)  \;  \left[ \pmatrix{1\cr 1\cr} \; + \; \Theta(\vec{x},t,p,\hat{p})  \right] \; \; ,
\end{equation}
where $ \Theta(\vec{x},t,p,\hat{p})$ is a two component column vector.  
Using Eq.~(\ref{1.3}) and defining 
\begin{equation}
\label{3.5}
\mu \;  = \;  \cos \theta_{\vec{p}  \vec{n}} \; \; , \; \;  \cos \theta_{\vec{k} \vec{ p}} \; = \;    \frac{\vec{k} \cdot \hat{p}}{k} \; ,
\end{equation}
we get from  Eq.~(\ref{3.3}):
\begin{eqnarray}
\label{3.6}
&& \frac{\partial \Theta(\vec{k},t,\mu)   }{\partial t}  \, + \,  \frac{i \, k}{a(t)}   \cos \theta_{\vec{k} \vec{ p}}   \; \Theta(\vec{k},t,\mu) \;
 \simeq  \; \frac{1}{2} \;  \left [ \frac{d}{d \, t} \, e^2(t) \right ] \;  \mu^2  \;  \pmatrix{1\cr 1\cr} \nonumber \\
&& - \, \sigma_T \, n_e \left [ \Theta(\vec{k},t, \mu) \,   - \,  \frac{3}{8} \int_{-1}^1
\pmatrix{2(1-\mu^2)(1-\mu^{\prime 2})+\mu^2 \mu^{\prime 2}&\mu^2\cr
\mu^{\prime 2}&1\cr} \,  \Theta(\vec{k},t, \mu') \; d\mu' \right] 
\end{eqnarray}
where $\sigma_T$ is the Thomson cross section and $ n_e(t)$  the electron number density ~\citep{Chandrasekhar:1960}. \\
Introducing the conformal time:
\begin{equation}
\label{3.7}
\eta(t) \; =  \;  \int_{0}^t \frac{dt'}{a(t')} \; \; ,
\end{equation}
we rewrite Eq.~(\ref{3.6}) as:
\begin{eqnarray}
\label{3.8}
&& \frac{\partial \Theta(\vec{k},\eta,\mu)   }{\partial \eta}  \, + \,  i \, k \;  \cos \theta_{\vec{k} \vec{ p}}   \; \Theta(\vec{k},\eta,\mu) \;
 \simeq  \; \frac{1}{2} \;  \left [ \frac{d}{d \, \eta} \, e^2(\eta) \right ] \; ( \mu^2 \, - \, \frac{1}{3})  \;  \pmatrix{1\cr 1\cr} \nonumber \\
&& -  a(\eta) \, \sigma_T \, n_e \left [ \Theta(\vec{k},\eta, \mu) \,   - \,  \frac{3}{8} \int_{-1}^1
\pmatrix{2(1-\mu^2)(1-\mu^{\prime 2})+\mu^2 \mu^{\prime 2}&\mu^2\cr
\mu^{\prime 2}&1\cr} \,  \Theta(\vec{k},\eta, \mu') \; d\mu' \right] 
\end{eqnarray}
with a suitable overall normalization of the blackbody intensity.
To determine the general solutions of Eq.~(\ref{3.8}) we write~\citep{Basko:1980,Cea:2010}:
\begin{equation}
\label{3.9}
\Theta(\vec{k},\eta,\mu)  \; =  \;   \theta_a(\vec{k},\eta) \, (\mu^2 - \frac{1}{3})   \pmatrix{1\cr 1\cr} \; +  \;   \theta_p(\vec{k},\eta) \, (1 - \mu^2 ) 
   \pmatrix{1\cr -1\cr} \; .
\end{equation}
From Eq.~(\ref{3.9}) it is evident that  $ \theta_a$ measures the degree of anisotropy, while $\theta_p$ gives the polarization of the primordial radiation.
With the aid of Eq.~(\ref{3.9}) we rewrite   Eq.~(\ref{3.8}) as:
\begin{eqnarray}
\label{3.10}
 && \frac{\partial \theta_a(\vec{k},\eta)}{\partial \eta}  \, + \,  i \, k \;  \cos \theta_{\vec{k} \vec{ p}}   \; \theta_a(\vec{k},\eta) \;
 \simeq  \; \Delta H(\eta)   \;  -  \;  a(\eta) \, \sigma_T \, n_e \left [  \frac{9}{10} \, \theta_a(\vec{k},\eta)  \; + \;   \frac{3}{5} \, \theta_p(\vec{k},\eta)  \right ]  
\nonumber \\
\\ 
&&  \frac{\partial \theta_p(\vec{k},\eta)}{\partial \eta}  \, + \,  i \, k \;  \cos \theta_{\vec{k} \vec{ p}}   \; \theta_p(\vec{k},\eta) \;
 \simeq  \;  -  \;  a(\eta) \, \sigma_T \, n_e  \left [ \frac{1}{10} \, \theta_a(\vec{k},\eta)  \; + \;   \frac{2}{5} \, \theta_p(\vec{k},\eta)  \right]  
  \nonumber  
\end{eqnarray}
where  we introduced the  cosmic shear~\citep{Negroponte:1980,Cea:2010}:
\begin{equation}
\label{3.11}
\Delta H(\eta) \;  \equiv  \;  \frac{1}{2} \; \frac{d }{d \eta}e^2(\eta)  \; \; .
\end{equation}
The  solution of the linear differential system Eq.~(\ref{3.10}) is the sum of the  general solution of the homogeneous system and a particular solution.
The solution of the homogeneous system (i.e. $\Delta H(\eta)=0$) is:
\begin{equation}
\label{3.12}
 \theta_a(\vec{k},\eta) \; = \;  \theta_p(\vec{k},\eta) \; = \; 0  \; ,
\end{equation}
for the are no anisotropies without cosmological perturbations. To determine the particular solution of  Eq.~(\ref{3.10}), we note that 
the linear combination:
\begin{equation}
\label{3.13}
\bar{ \theta}(\vec{k},\eta)   \; \equiv \;  \theta_a(\vec{k},\eta)  \; + \;   \theta_p(\vec{k},\eta)  
\end{equation}
satisfies the following equation:
\begin{equation}
\label{3.14}
 \frac{\partial \bar{\theta}(\vec{k},\eta)}{\partial \eta}  \, + \,  i \, k \;  \cos \theta_{\vec{k} \vec{ p}}   \; \bar{\theta}(\vec{k},\eta) \;
 \simeq  \; \Delta H(\eta)   \;  -  \;  a(\eta) \, \sigma_T \, n_e \,  \bar{\theta}(\vec{k},\eta) \; .
\end{equation}
Introducing the optical depth:
\begin{equation}
\label{3.15}
 \tau(\eta,\eta') =   \int_{\eta'}^{\eta} \sigma_T \, n_e \, a(\eta'') \, d\eta'' \; \; ,
\end{equation}
 it is easy to verify that the solution of Eq.~(\ref{3.14}) is given by:
\begin{equation}
\label{3.16}
 \bar{\theta}(\vec{k},\eta) =  \int_{\eta_i}^{\eta} \Delta H(\eta')  \;   e^{-\tau(\eta,\eta')} \;    e^{i \, k \,   \cos \theta_{\vec{k} \vec{ p}} (\eta' - \eta)}  \, d\eta'  \; ,
\end{equation}
where $\eta_i$ is an early conformal time such that  $\bar{\theta}(\vec{k},\eta_i)=0$. It is now easy to determine $\theta_a$ and $\theta_p$.
We get:
\begin{equation}
\label{3.17}
  \theta_a(\vec{k},\eta) = \frac{1}{7}  \int_{\eta_i}^{\eta} \Delta H(\eta') \left[ 6 e^{-\tau(\eta,\eta')} \,  +  \, e^{- \frac{3}{10} \tau(\eta,\eta')}  \right] 
  \;  e^{i \, k \,   \cos \theta_{\vec{k} \vec{ p}} (\eta' - \eta)}  \;  d\eta'   \; ,
\end{equation}
\begin{equation}
\label{3.18}
  \theta_p(\vec{k},\eta) = \frac{1}{7}  \int_{\eta_i}^{\eta} \Delta H(\eta') \left[  e^{-\tau(\eta,\eta')} \,  - \, e^{- \frac{3}{10} \tau(\eta,\eta')}  \right] 
  \;  e^{i \, k \,   \cos \theta_{\vec{k} \vec{ p}} (\eta' - \eta)}  \;  d\eta'   \; .
\end{equation}
In summary, we have found that  the temperature fluctuations  induced by the spatial anisotropy of the geometry of the universe 
at large scales is given by Eqs.~(\ref{3.9}),  (\ref{3.17})  and (\ref{3.18}).  Obviously, we are interested in the temperature
anisotropies for $\eta=\eta_0$  ($\eta_0$ is the conformal time now). As will be evident later on, the main contributions
to the integrals in  Eqs.~(\ref{3.17})  and (\ref{3.18}) come from conformal times near the decoupling conformal time $\eta_d$.
Moreover, observing that $\eta_d \ll \eta_0$ we may write:
\begin{equation}
\label{3.19}
\Theta(\vec{k},\eta_0,\mu, \hat{p})  \; \simeq \;   \theta_a \, (\mu^2 - \frac{1}{3})  \;  e^{- \, i \, k \,   \cos \theta_{\vec{k} \vec{ p}} \,  \eta_0} \;  \pmatrix{1\cr 1\cr} \; +  \;  
 \theta_p \, (1 - \mu^2 ) \;  e^{- \, i \, k \,   \cos \theta_{\vec{k} \vec{ p}} \,  \eta_0} \;    \pmatrix{1\cr -1\cr} \; ,
\end{equation}
where:
\begin{equation}
\label{3.20}
  \theta_a \; \simeq \;  \frac{1}{7}  \int_{\eta_i}^{\eta_0} \Delta H(\eta') \left[ 6 e^{-\tau(\eta_0,\eta')} \,  +  \, e^{- \frac{3}{10} \tau(\eta_0,\eta')}  \right]  \;  d\eta'   \; ,
\end{equation}
\begin{equation}
\label{3.21}
  \theta_p \;  \simeq \;  \frac{1}{7}  \int_{\eta_i}^{\eta_0} \Delta H(\eta') \left[  e^{-\tau(\eta_0,\eta')} \,  - \, e^{- \frac{3}{10} \tau(\eta_0,\eta')}  \right]   \;  d\eta'   \; .
\end{equation}
In appendix~\ref{Appendix A} we evaluate the two parameters $\theta_a$ and $\theta_p$. We find (see  Eqs.~(\ref{A.15})  and (\ref{A.8}) ):
\begin{equation}
\label{3.20-bis}
\theta_a \;  \simeq \;  - \; \frac{1}{2}   \times 0.944  \;  e^2_{\rm dec}  \; ,
\end{equation}
\begin{equation}
\label{3.21-bis}
\theta_p \;  \simeq \;  8.92 \; 10^{-3} \;  e^2_{\rm dec}  \;   \; .
\end{equation}
\section[]{The quadrupole anomaly}
\label{S4}
We are, now, in position to discuss the low quadrupole anomaly in the CMB temperature anisotropies detected by WMAP and recently confirmed by Planck.
The temperature anisotropies of the cosmic background depend on the polar angle $\theta, \phi$, so that one usually expands in terms of spherical harmonics:
\begin{equation}
\label{4.1}
\frac{\Delta T(\theta,\phi)}{ T_0} \;  = \; 
\sum_{\ell = 1 }^{\infty} \sum_{m= - \ell}^{+ \ell} \; a_{\ell m} \, Y_{\ell m}(\theta,\phi)  \; ,
\end{equation}
where  $T_0 \simeq 2.7255 \; K$~\citep{Fixsen:2009} is the actual (average) temperature of the CMB radiation.
Note that the $a_{\ell m}$'s  in  Eq.~(\ref{4.1}) are dimensionless and are obtained from
the corresponding coefficients in  Eq.~(\ref{1.5})  by dividing by $T_0$. 
After that, one introduces the power spectrum:
\begin{equation}
\label{4.2}
( \frac{\Delta T_{\ell}}{ T_0 } )^2 \; =  \;  \frac{1}{2 \pi} \,
\frac{\ell (\ell+1)}{2 \ell + 1} \sum_m | a_{\ell m} |^2 \; ,
\end{equation}
that fully characterizes the properties of the CMB temperature anisotropy. In
particular, we focus on  the quadrupole anisotropy $\ell=2$:
\begin{equation}
\label{4.3}
\mathcal{Q}^2  \;  \equiv  \, ( \frac{\Delta T_2}{ T_0} )^2  \;  .
\end{equation}
In the standard model the CMB temperature fluctuations are induced by the cosmological perturbations of
the  FRW homogeneous and isotropic background metric generated by the inflation-produced potentials.
In the ellipsoidal universe we must also consider the effects on the CMB anisotropies induced
by the anisotropic expansion of the universe. In fact, as discussed in sec.~\ref{S2}, at large scales the observed
anisotropies in the CMB temperature are due to the linear superposition of the two contributions according
to  Eq.~(\ref{2.32}).  Therefore, we may write:
\begin{equation}
\label{4.4}
a_{\ell m} \; =  \; a_{\ell m}^{ A} \;  + \; a^{ I}_{\ell m} \; .
\end{equation}
In the previous section we have determined the contributions to  temperature contrast function induced
by the   anisotropic expansion of the universe:
\begin{equation}
\label{4.5}
\Theta^{A}(\vec{k},\eta_0,\mu, \hat{p})  \; \simeq \;   \theta_a \, (\mu^2 - \frac{1}{3})  \;  e^{- \, i \, k \,   \cos \theta_{\vec{k} \vec{ p}} \,  \eta_0}  \; \; , \; \;   
\mu \;  = \;  \cos \theta_{\vec{p}  \vec{n}} \; .
\end{equation}
In appendix~\ref{Appendix B}, starting from  Eq.~(\ref{4.5})   we perform the multipole expansion of the temperature fluctuation correlations
and obtain the multipole coefficients  $a_{\ell m}^{ A}$. However, it is evident from  Eq.~(\ref{4.5})   that the main contribution
to the temperature fluctuations is for $k \, \simeq \, 0$. It is easy to see that this corresponds to solve the Boltzmann
equation Eq.~(\ref{3.1})  by neglecting the spatial dependence on the temperature contrast function.  In this case we
obtain at once:
\begin{equation}
\label{4.6}
\frac{\Delta T^{A}(\theta,\phi)}{ T_0} \;  \simeq \;  \theta_a \, ( \cos^2 \theta_{\vec{p}  \vec{n}}  - \frac{1}{3})  \;  ,    
\end{equation}
where $\theta_a$ is given by  Eq.~(\ref{3.20-bis}) and $\theta,\phi$ are the polar angles of the photon momentum $\vec{p}$. \\
Let $\theta_n, \phi_n$ be the polar angles of the direction of the axis of symmetry $\vec{n}$, then:
\begin{equation}
\label{4.7}
\frac{\Delta T^{A}(\theta,\phi)}{ T_0} \;  \simeq \; \frac{2}{3} \;  \theta_a \; P_2( \cos \theta_{\vec{p}  \vec{n}})   \;  
= \; \frac{2}{3} \;  \theta_a \;  \frac{ 4 \pi}{5} \, \sum_{m= - 2}^{+ 2} \;  \, Y_{2 m}(\theta,\phi)  \, Y^*_{2 m}(\theta_n,\phi_n)  \;  .    
\end{equation}
Since from Eq.~(\ref{4.1}) it follows that:
\begin{equation}
\label{4.8}
a_{\ell m}^{ A}  \; = \; \int \, d\Omega \;  \frac{\Delta T^{A}(\theta,\phi)}{ T_0} \;  Y^*_{\ell m}(\theta,\phi)  \;  ,    
\end{equation}
we obtain immediately:
\begin{equation}
\label{4.9}
a_{2 m}^{ A}  \; \simeq \;   - \,  \frac{ 4 \pi}{15}  \, \epsilon^2   \;  Y^*_{\ell m}(\theta_n,\phi_n)  \;  \; , \; \; \epsilon^2 \; \equiv \;   0.944  \;  e^2_{\rm dec}  \; ,
\end{equation}
while $a_{\ell m}^{ A} = 0$ for $\ell \neq 2$.  In other words, at large scales the anisotropy of the metric contributes mainly to the quadrupole CMB temperature anisotropies.
From Eq.~(\ref{4.9}) we find:
\begin{eqnarray}
\label{4.10}
&& a_{20}^{A} \; \simeq \;  - \, \frac{ 4 \pi}{15}  \, \epsilon^2   \sqrt{ \frac{5}{16 \pi} } \,
                  [1 - 3 \cos^2 \theta_n ] \;  ,  \nonumber \\
&& a_{21}^{ A} \;  = \;  (a_{2,-1}^{A})^{*} \; \simeq \; + \, i \, 
                   \frac{ 4 \pi}{15}  \, \epsilon^2   \sqrt{ \frac{15}{8 \pi} }
                            e^{-i \phi_n}  \sin \theta_n  \, \cos \theta_n  \;  , \nonumber \\
&& a_{22}^{ A} \; = \; (a_{2,-2}^{ A})^{*} \;  \simeq \;
                    \frac{ 4 \pi}{15}  \, \epsilon^2   \sqrt{ \frac{15}{32 \pi} }
                  \; e^{-2 i \phi_n}  \sin^2 \theta_n  \; .
\end{eqnarray}
After a little algebra we rewrite Eq.~(\ref{4.10}) as:
\begin{eqnarray}
\label{4.11}
&& a_{20}^{A} \; \simeq \;  + \, \frac{ 1}{6}  \, \epsilon^2   \sqrt{ \frac{\pi}{5} } \,
                  [1 + 3 \cos^2 (2 \theta_n) ] \;  ,  \nonumber \\
&& a_{21}^{ A} \;  = \;  (a_{2,-1}^{A})^{*} \; \simeq \; + \, i \, 
                  \sqrt{  \frac{\pi}{30} }  \, \epsilon^2 
                            e^{-i \phi_n}  \sin (2 \theta_n )  \;  , \nonumber \\
&& a_{22}^{ A} \; = \; (a_{2,-2}^{ A})^{*} \;  \simeq \; + \,
                   \sqrt{  \frac{ \pi}{30} } \, \epsilon^2   
                  \; e^{-2 i \phi_n}  \sin^2 \theta_n  \; .
\end{eqnarray}
Defining the quadrupole anisotropy:
\begin{equation}
\label{4.12}
\mathcal{Q}_A^2  \;  \equiv  \, ( \frac{\Delta T^A_2}{ T_0} )^2  \;  ,
\end{equation}
we find:
\begin{equation}
\label{4.13}
\mathcal{Q}_{ A} \;  \simeq \; \frac{2}{5 \sqrt{3} } \; \epsilon^2  \;  .
\end{equation}
To determine the coefficients $a_{\ell m}$, Eq.~(\ref{4.4}), we need to known the $a^I_{\ell m}$'s. First we observe that 
 the temperature anisotropies are real functions, so that we must have $a_{\ell,-m} = (-1)^m (a_{\ell,m})^*$. 
 Observing that $a^A_{\ell,-m} = (-1)^m (a^A_{\ell,m})^*$ (see Eq.~(\ref{4.11})), we have the constraints
$a^{I}_{\ell,-m} = (-1)^m (a^{I}_{\ell,m})^*$.
Moreover, because the standard inflation-produced temperature
fluctuations are statistically isotropic, we reasonably  assume that the $a^{ I}_{2m}$ coefficients are
equals up to a phase factor. Therefore, we can write:
\begin{eqnarray}
\label{4.14}
&& a^{I}_{20} \; \simeq \;  \sqrt{\frac{\pi}{3}} \;   \mathcal{Q}_{I}, \nonumber  \\
&& a^{I}_{21} \; = \;  - \,  (a^{\rm I}_{2,-1})^{*} \; \simeq \; + \, i \,  \sqrt{\frac{\pi}{3}} \; e^{i \phi_1} \;  \mathcal{Q}_{ I}  \; , \\
&& a^{ I}_{22} \; =  \; (a^{\rm I}_{2,-2})^{*} \;  \simeq \; \sqrt{\frac{\pi}{3}}
\; e^{i \phi_2}  \; \mathcal{Q}_{I}  \;  , \nonumber
\end{eqnarray}
where $0 \leq \phi_1, \phi_2  \leq 2 \pi$ are unknown phases. It is easy to check that:
\begin{equation}
\label{4.15}
\mathcal{Q}_I^2  \;  =  \;  ( \frac{\Delta T^I_2}{ T_0} )^2  \;  .
\end{equation}
Using  Eq.~(\ref{1.7}) we obtain the estimate:
\begin{equation}
\label{4.16}
\mathcal{Q}_I  \;  \simeq  \; ( \, 12.44 \; \pm \; 3.93 \, ) \; 10^{-6}  \;  .
\end{equation}
Taking into account Eqs.~(\ref{4.3}),  (\ref{4.4}),   (\ref{4.11}) and (\ref{4.14})  we get for the total quadrupole:
\begin{equation}
\label{4.17}
\mathcal{Q}^2 = \mathcal{Q}_{A}^2 \; + \; \mathcal{Q}_{I}^2 \;  + \;  2 \, f(\theta_n,\phi_n,\phi_1,\phi_2) \,
\mathcal{Q}_{A} \mathcal{Q}_{I} \; ,
\end{equation}
where
\begin{equation}
\label{4.18}
f(\theta_n,\phi_n,\phi_1,\phi_2)  \;  = \;  
             \frac{1}{4\sqrt{5}} \, [1 + 3 \cos(2 \theta_n) ]  \; + \;
                        \sqrt{ \frac{3}{10} } \, \sin (2 \theta_n) \, \cos (\phi_1 + \phi_n)  \; + \; 
               \sqrt{ \frac{3}{10} } \,  \sin^2 \theta_n \, 
               \cos ( \phi_2 + 2 \phi_n ) \; .
\end{equation}
Eqs.~(\ref{4.17}) and  (\ref{4.18}) show that, indeed, if the space-time background metric is not isotropic, the quadruple anisotropy
may become smaller than the one expected in the standard isotropic $\Lambda$CDM cosmological model of temperature
fluctuations. In fact, from Eq.~(\ref{4.4}) and using  Eqs.~(\ref{4.11}) and  (\ref{4.14}), we get:
\begin{equation}
\label{4.19}
 a_{20}  \; \simeq \;  + \,    \sqrt{\frac{\pi}{3}} \;   \mathcal{Q}_{I} \; + \; \frac{ 1}{6}  \, \epsilon^2   \sqrt{ \frac{\pi}{5} } \,
                  [1 + 3 \cos^2 (2 \theta_n) ]    \;  ,  
\end{equation}
\begin{equation}
\label{4.20}
 a_{21}  \;  = \;   + \, i \,  \sqrt{\frac{\pi}{3}} \; e^{i \phi_1} \;  \mathcal{Q}_{ I}  \; + \;
                      \, i \, \sqrt{ \frac{\pi}{30}}  \, \epsilon^2  \, e^{-i \phi_n}  \sin (2 \theta_n )  \;  , 
\end{equation}
\begin{equation}
\label{4.21}
a_{22}  \;    \simeq  \;  + \, \sqrt{\frac{\pi}{3}} \; e^{i \phi_2}  \; \mathcal{Q}_{I}  \;  + \; 
                                              \sqrt{  \frac{ \pi}{30} }  \, \epsilon^2   \; e^{-2 i \phi_n}  \sin^2 \theta_n   \; .
\end{equation}
Eqs.~(\ref{4.19}),  (\ref{4.20})  and  (\ref{4.21}) give a system of five equations which can be solved to get the five unknown
parameters $e^2_{\rm dec}, \theta_n, \phi_n, \phi_1, \phi_2$. To do this, however, we need the observed values of the $a_{\ell,m}$'s. In fact,
\citet{Campanelli:2007} used the cleaned  CMB temperature fluctuation maps of the WMAP data obtained 
using the internal linear combination with galactic foreground subtraction. In particular, these authors used
three different maps~\citep{Hinshaw:2007,Oliveira:2006,Park:2007}.  
Actually, the same procedure can be applied to the foreground-cleaned CMB maps obtained from the Planck data as detailed in \citet{Ade:2013b}.
However, irrespective from the adopted CMB cleaned map the quadrupole anomalies detected by WMAP and confirmed by Planck
are accounted for if:
\begin{equation}
\label{4.22}
 a_{21}  \;  \approx \;   0  \;  , 
\end{equation}
and
\begin{equation}
\label{4.23}
 |a_{20}|^2 \; \ll   2 \, | a_{22} |^2   \;  . 
\end{equation}
In fact, it is easy to check that these equations imply both the almost planarity and the suppression of power
of the quadrupole moment. Remarkably, it turns out that  Eqs.~(\ref{4.22})  and  (\ref{4.23})  allow us to determine
the eccentricity at decoupling and constraint the polar angles of the symmetry axis.
Inserting Eq.~(\ref{4.22}) into Eq.~(\ref{4.20}) we readily obtain:
\begin{equation}
\label{4.24}
 \epsilon^2  \; \simeq \;  \frac{ \sqrt{10} \, \mathcal{Q}_{ I}}{| \sin (2 \theta_n )|}   \;  , 
\end{equation}
where $\phi_n + \phi_1 \simeq 0^{\circ} \, , \, 360^{\circ}$ if $\sin (2 \theta_n ) < 0$, or  $\phi_n + \phi_1 \simeq 180^{\circ} \, , \, 540^{\circ}$ if $\sin (2 \theta_n ) > 0$.
Moreover, from  Eq.~(\ref{4.19}) and taking into account  Eq.~(\ref{4.23}) we find:
\begin{equation}
\label{4.25}
\cos (2 \theta_n ) \; \simeq \;  - \, \frac{1}{3} \; - \, 2 \,  \sqrt{ \frac{5}{3}} \, \frac{ \mathcal{Q}_{ I} }{\epsilon^2}  \; . 
\end{equation}
Combining Eqs.~(\ref{4.24})  and  (\ref{4.25}) we obtain:
\begin{equation}
\label{4.26}
 \theta_n  \; \simeq \;   \arctan \left ( \pm \frac{\sqrt{6}}{2} \; + \; 2 \right )  \; \simeq  \;   73^{\circ} \, , \, 107^{\circ} \; .
\end{equation}
This last equation together with Eqs.~(\ref{4.16}) and Eq.~(\ref{4.24}) gives the eccentricity at decoupling: 
\begin{equation}
\label{4.27}
 e_{\rm dec} \;  \simeq  \;   ( 0.86 \, \pm \, 0.14) \, 10^{-2}   \; .
\end{equation}
Finally, using   Eqs.~(\ref{4.3}), (\ref{4.22}) and   (\ref{4.23})  we get:
\begin{equation}
\label{4.28}
 \mathcal{Q}^2 \; \simeq \;  \frac{6}{5 \pi} \;  | a_{22} |^2 \; \simeq \;   \frac{6}{25 \pi} \,   \mathcal{Q}^2_{I}  
\left [ 1 \, + \, \frac{  \sin^4 \theta_n}{  \sin^2 (2 \theta_n )} \, +  \, \frac{ 2  \sin^2 \theta_n}{  |\sin (2 \theta_n )|} \,   \cos (\phi_2 + 2 \phi_n) \right ] \; .
\end{equation}
Using the observed value of the quadrupole temperature anisotropy    Eq.~(\ref{1.6}) we estimate from Eq.~(\ref{4.28}) :
\begin{equation}
\label{4.29}
\cos (\phi_2 + 2 \phi_n) \;  \simeq  \;   - \;  0.92  \;  \pm \;  0.12     \; .
\end{equation}
To summarize, our almost model independent analysis allowed to fix the eccentricity at decoupling,  Eq.~(\ref{4.27}). As concern the symmetry axis,
using the galactic coordinates $b_n,l_n$,  we found:
\begin{equation}
\label{4.30}
b_n \;  \simeq  \;   \pm  \,  17^{\circ}  \; ,
\end{equation}
while the longitude $l_n$ turned  out to be poorly constrained in qualitative agreement with  \citet{Campanelli:2007}.
\section{The large scale polarization}
\label{S5}
In this section we discuss the large scale polarization in the primordial cosmic background. In our previous work~\citep{Cea:2010} we argued that
the ellipsoidal geometry of the universe induces sizable polarization signal at large scale without invoking the CMB reionization mechanism.
If we assume that early CMB reionization is negligible, then it is well known that at large scale the primordial inflation induced cosmological
perturbations do not produce sizable polarization signal~\citep{Dodelson:2003,Mukhanov:2005}. In this case the polarization of the temperature fluctuations
are fully given by the anisotropic expansion of the universe. According to our discussion in sect.~\ref{S3} we may write:
\begin{equation}
\label{5.1}
\Theta^{E}(\vec{k},\eta_0,\mu, \hat{p})  \; \simeq \;   \theta_p \, (1 \, - \, \cos^2 \theta_{\vec{p}  \vec{n}} )  \;  e^{- \, i \, k \,   \cos \theta_{\vec{k} \vec{ p}} \,  \eta_0}  \; ,
\end{equation}
where the superscript E  indicates that the temperature polarization contributes only to the so-called E-modes. In fact,  Eq.~(\ref{3.19}) shows that the 
anisotropy of the metric of the universe gives rise only to a linear polarization of the cosmic background radiation. 
In appendix~\ref{Appendix B}   we discuss the multipole expansion of the temperature polarization correlations. 
As we have already observed,   the main contribution to the polarization temperature contrast functions  is for $k \, \simeq \, 0$, which
 corresponds to neglect  the spatial dependence of the solutions of the Boltzmann
equation. Thus, we have:
\begin{equation}
\label{5.2}
\frac{\Delta T^{E}(\theta,\phi)}{ T_0} \;  \simeq \;  \theta_p \, (1 \, - \, \cos^2 \theta_{\vec{p}  \vec{n}} ) \; = \;
 \frac{2}{3} \, \theta_p  \; - \; \frac{2}{3} \;  \theta_p \; P_2( \cos \theta_{\vec{p}  \vec{n}})   \; .  
\end{equation}
We may, now,   expand in terms of spherical harmonics as in Eq.~(\ref{4.1}).  It is evident from  Eq.~(\ref{5.2}) that  
the non-zero multipole coefficients  $a_{\ell m}^{ E}$ are for the monopole $\ell = 0$ and the quadrupole $\ell = 2$.
The monopole term determines the average large scale  polarization of the cosmic microwave background:
\begin{equation}
\label{5.3}
\frac{ \Delta T_{pol}}{T_0} \; \equiv \; \frac{1}{4 \pi} \; \int \; d\Omega \; \;  \frac{\Delta T^{E}(\theta,\phi)}{ T_0} \;  \simeq \;  
 \frac{2}{3} \, \theta_p   \; .
\end{equation}
Using  Eqs.~(\ref{3.21-bis}) and  (\ref{4.27}) we obtain:
\begin{equation}
\label{5.4}
 \Delta T_{pol} \;  \simeq  \;  (1.20  \, \pm  \,  0.38) \;  \mu K  \; ,
\end{equation}
in qualitative agreement with our previous estimate~\citep{Cea:2010}. \\
On the other hand, from Eq.~(\ref{5.2}) we easily obtain:
\begin{equation}
\label{5.5}
a_{2 m}^{ E}  \; \simeq \;   - \,  \frac{ 8 \pi}{15}  \, \theta_p   \;  Y^*_{2 m}(\theta_n,\phi_n)  \;  \; , 
\end{equation}
which implies:
\begin{eqnarray}
\label{5.6}
&& a_{20}^{E} \; \simeq \;  - \, \frac{ 8 \pi}{15}  \,  \theta_p  \,  \sqrt{ \frac{5}{16 \pi} } \,
                  [1 - 3 \cos^2 \theta_n ] \;  ,  \nonumber \\
&& a_{21}^{ E} \;  = \;  (a_{2,-1}^{A})^{*} \; \simeq \; + \, i \, 
                   \frac{ 8 \pi}{15}  \,  \theta_p  \,  \sqrt{ \frac{15}{8 \pi} }
                            e^{-i \phi_n}  \sin \theta_n  \, \cos \theta_n  \;  , \nonumber \\
&& a_{22}^{ E} \; = \; (a_{2,-2}^{ A})^{*} \;  \simeq \;
                    \frac{ 8 \pi}{15}  \, \theta_p \,  \sqrt{ \frac{15}{32 \pi} }
                  \; e^{-2 i \phi_n}  \sin^2 \theta_n  \; ,
\end{eqnarray}
or better:
\begin{eqnarray}
\label{5.7}
&& a_{20}^{E} \; \simeq \;  + \, \frac{ 1}{3}  \,  \theta_p  \,    \sqrt{ \frac{\pi}{5} } \;
                  [1 + 3 \cos^2 (2 \theta_n) ] \;  ,  \nonumber \\
&& a_{21}^{ E} \;  = \;  (a_{2,-1}^{A})^{*} \; \simeq \; + \, 2 \; i \;
                  \sqrt{  \frac{\pi}{30} }  \; \theta_p  \, 
                            e^{-i \phi_n}  \sin (2 \theta_n )  \;  , \nonumber \\
&& a_{22}^{ E} \; = \; (a_{2,-2}^{ A})^{*} \;  \simeq \; + \, 2 \,
                   \sqrt{  \frac{ \pi}{30} } \;  \theta_p  \,   
                  \; e^{-2 i \phi_n}  \sin^2 \theta_n  \; .
\end{eqnarray}
Eq.~(\ref{5.7}) allows to evaluate the quadrupole EE correlation:
\begin{equation}
\label{5.8}
( \frac{\Delta T^{EE}_{2}}{ T_0 } )^2 \; =  \;  \frac{3}{5 \,  \pi} \;
 \sum_{m=-2}^{m=+2}  | a^{E}_{2 m} |^2 \;  \; . 
\end{equation}
In fact, a straightforward calculation gives:
\begin{equation}
\label{5.9}
( \frac{\Delta T^{EE}_{2}}{ T_0 } )^2  \;  \simeq \; \frac{16}{75} \;   \theta_p^2  \; . 
\end{equation}
Using again  Eqs.~(\ref{3.21-bis}) and  (\ref{4.27}) we find:
\begin{equation}
\label{5.10}
\Delta T^{EE}_{2}   \;  \simeq \;  0.83 \; \pm \; 0.27  \;   \mu \, K \; .
\end{equation}
We may, also, estimate the quadrupole TE correlation:
\begin{equation}
\label{5.11}
( \frac{\Delta T^{TE}_{2}}{ T_0 } )^2 \; =  \;  \frac{3}{5 \,  \pi} \;
 \sum_{m=-2}^{m=+2} \,   a^{T}_{2 m} \,  (a^{E}_{2 m})^*  \;  =  \;  \frac{3}{5 \,  \pi} \; \left \{ a^{T}_{2 0} \,  a^{E}_{2 0}  \, + \,
2 \,  \mathcal{R}e \,  [ a^{T}_{2 1} \,  (a^{E}_{2 1})^*]  \, + \,
2 \,  \mathcal{R}e \,  [ a^{T}_{2 2} \,  (a^{E}_{2 2})^*]  \right \} \; ,
\end{equation}
where the $ a^{T}_{2 m}$'s are given by Eqs.~(\ref{4.19}) -  (\ref{4.21}).  Using   Eqs.~(\ref{4.22}) and   (\ref{4.23}),
we may simplify    Eq.~(\ref{5.11}) as:
\begin{equation}
\label{5.12}
( \frac{\Delta T^{TE}_{2}}{ T_0 } )^2  \;  \simeq  \;  \frac{6}{5 \,  \pi} \; 
 \mathcal{R}e \,  [ a^{T}_{2 2} \,  (a^{E}_{2 2})^*]   \; \simeq \;  \frac{4 }{5 \,  \sqrt{10}} \; \theta_p \;  \sin^2 \theta_n \,
 \left [    \mathcal{Q}_I \, \cos (\phi_2 + 2 \, \phi_n) \; + \; \frac{1}{\sqrt{10}} \, \epsilon^2 \sin^2 \theta_n \right ] \;.
\end{equation}
After using Eq.~(\ref{4.24}), this last equation can be rewritten as:
\begin{equation}
\label{5.13}
( \frac{\Delta T^{TE}_{2}}{ T_0 } )^2  \;  \simeq  \; \frac{4 }{5 \,  \sqrt{10}} \; \theta_p \; \mathcal{Q}_I  \;  \sin^2 \theta_n \;
 \left [ \cos (\phi_2 + 2 \, \phi_n) \; + \; \frac{  \sin^2 \theta_n }{ | \sin (2 \, \theta_n) |}   \right ] \;.
\end{equation}
Finally, using our previous estimates  Eqs.~(\ref{4.26}) and (\ref{4.29}) we find:
\begin{equation}
\label{5.14}
\Delta T^{TE}_{2}   \;  \simeq  \;  3.14  \; \pm  \; 0.76   \;   \mu \, K \; .
\end{equation}
\section{Conclusions}
\label{S6}
In this paper we solved at large scales the Boltzmann equation for the CMB photon distribution function by considering
the effects of the inflation primordial scalar perturbations and the anisotropy of the geometry in the ellipsoidal universe model.
We showed explicitly that the CMB temperature fluctuations are obtained by the linear superimposition  of the
temperature fluctuations induced by the cosmological scalar perturbations and by the spatial anisotropy of the metric.
We found that the anisotropic expansion of the universe, the so-called cosmic shear, affects mainly  the quadrupole
correlation functions. Moreover, we showed that these effects extend also to the low-lying multipoles $\ell \; \sim \; 10$. \\ 
We confirmed previous results that the low quadrupole temperature correlation, detected by WMAP and by the Planck satellite, could be 
accounted for if the geometry of the universe is plane-symmetric with  eccentricity  at decoupling of order $10^{-2}$.
We showed that the ellipsoidal geometry  of the universe produces sizable polarization signal at large scales. We found that 
our estimate of the quadrupole TE correlation were in agreement  both in sign and magnitude with observations. On the other hand, regarding the quadrupole
EE correlation our result did not compare well with the  final analysis of the Wilkinson Microwave
Anisotropy Probe collaboration.  However, we feel that the rather low polarization signal detected by WMAP at large scales could be due to an overestimation of the foreground
polarization signal. In fact, in the standard reionization scenario the large scale polarization in the temperature fluctuations is produced by the fraction
of the rescattered photons on the scales corresponding to the reionization horizon. As a consequence in this usually adopted scenario the polarization 
anisotropies are present for $\ell \ge 2$. That means, in particular, that  there is no average polarization. On the other hand, we have shown that
the anisotropic expansion in the ellipsoidal universe model implies the presence of  large scale polarization in the temperature fluctuations
without invoking  reionization processes. In fact, at variance with the usually accepted scenario, in the ellipsoidal universe we have
a sizeable average polarization signal at level $\sim \; \mu K$.  If this average polarization in the temperature fluctuations
of the cosmic background is misinterpreted as foreground polarization signal, then it could result in  a considerable underestimate
of the CMB polarization signal at large scales. Therefore  a careful characterization  of foreground polarization is certainly crucial for polarization measurements. \\
\indent
In conclusion, we are reinforcing  the proposal  that the ellipsoidal universe cosmological model is a viable alternative that
could  account for the detected large scale anomalies in the cosmic microwave anisotropies.
\appendix

\section{Evaluation of the parameters $\theta_{a}$ and $\theta_{p}$}
\label{Appendix A}
In this Appendix we evaluate the parameters $\theta_a$ and $\theta_p$  given by  Eqs.~(\ref{3.20}) and  (\ref{3.21}).  In fact,
these two parameters have been estimate in \citet{Cea:2010} by assuming that the plane-symmetric geometry is  induced by a cosmological magnetic field.
Presently, we would like to present a slightly better  estimate which is valid irrespective of the physical mechanism responsible for the
 generation of the spatial anisotropy in the early universe. \\
Let us consider, firstly, the parameter $\theta_p$. Defining $\tau(\eta) =  \tau(\eta_0, \eta)$, it is easy to verify that  $\tau(\eta', \eta) = \tau(\eta') - \tau(\eta)$. 
Observing that  $\tau(\eta_0) \simeq 0$, we rewrite  Eq.~(\ref{3.21}) as:
\begin{equation}
\label{A.1}
  \theta_p \;  \simeq \;  \frac{1}{7}  \int_{\eta_i}^{\eta_0} \Delta H(\eta') \left[  e^{-\tau(\eta')} \,  - \, e^{- \frac{3}{10} \tau(\eta')}  \right]   \;  d\eta'   \; .
\end{equation}
It is convenient to rewrite the integral in  Eq.~(\ref{A.1}) in terms of the cosmic time:
\begin{equation}
\label{A.2}
  \theta_p \;  \simeq \;  \frac{1}{7}  \int_{t_i}^{t_0}  \frac{1}{2} \; \frac{d }{d t'}e^2(t')  \left[  e^{-\tau(t')} \,  - \, e^{- \frac{3}{10} \tau(t')}  \right]   \;  d t'   \; ,
\end{equation}
where $t_0$ is the age of the universe and we used Eq.~(\ref{3.11}). To evaluate the derivative in  Eq.~(\ref{A.2}), we note that  
in general~\citep{Campanelli:2007}  $e^2(t) \sim a(t)^{-\frac{3}{2}}$.  Thus, in the matter-dominated era   we may write near decoupling:
\begin{equation}
\label{A.3}
 \frac{1}{2} \; \frac{d }{d t}e^2(t) \; \simeq \; - \; \frac{3}{4} \;  e^2(t) \; H(t) \; .
\end{equation}
After changing the integration variable by using instead of the cosmic time t the red-shift z, we obtain:
\begin{equation}
\label{A.4}
  \theta_p \;  \simeq \;  - \; \frac{3}{28}  \int_{0}^{\infty}  \; \frac{e^2(z') }{1 + z'}   \left[  e^{-\tau(z')} \,  - \, e^{- \frac{3}{10} \tau(z')}  \right]   \;  d z'   \; .
\end{equation}
Since near decoupling we may write:
\begin{equation}
\label{A.5}
 e^2(z) \;  \simeq \;  e^2_{\rm dec}  \;  (\frac{1 \, + \, z}{1 \, + z_d} )^{\frac{3}{2}}    \; \; , \; \; e^2_{\rm dec}  \;  =  \; e^2(z_d)  \; ,
\end{equation}
where $z_d \simeq 1090$ is the red-shift at decoupling, we get:
\begin{equation}
\label{A.6}
  \theta_p \;  \simeq \;  - \; \frac{3}{28}  \, e^2_{\rm dec} \; \int_{0}^{\infty}  \; (\frac{1 \, + \, z}{1 \, + z_d} )^{\frac{3}{2}}   
  \frac{1 }{1 + z}   \left[  e^{-\tau(z)} \,  - \, e^{- \frac{3}{10} \tau(z)}  \right]   \;  d z   \; .
\end{equation}
Finally, it is known that near decoupling to a good approximation one can write~\citep{Jones:1985}:
\begin{equation}
\label{A.7}
 \tau(z) \;  \simeq \;  0.37  \;  (\frac{ z}{1000} )^{14.25}    \; \; , \; \; 500 \;  \lesssim  \; z \;  \lesssim  \;  1400   \; \; .
\end{equation}
This allow us to evaluate numerically the integral in Eq.~(\ref{A.6}). We obtain:
\begin{equation}
\label{A.8}
\theta_p \;  \simeq \;  8.92 \; 10^{-3} \;  e^2_{\rm dec}  \;   \; .
\end{equation}
To evaluate the parameter  $\theta_a$, we note that:
\begin{equation}
\label{A.9}
  \theta_a \; \simeq \;  \frac{1}{7}  \int_{\eta_i}^{\eta^*} \Delta H(\eta') \left[ 6 e^{-\tau(\eta')} \,  +  \, e^{- \frac{3}{10} \tau(\eta')}  \right]  \;  d\eta'   \; 
  + \;  \int_{\eta^*}^{\eta_0} \; \Delta H(\eta') \;  d\eta'  \;  ,
\end{equation}
where $\eta^*$  is a conformal time such that  $\tau(\eta) = 0$ for $\eta \ge \eta^*$. The second integral in the right hand in Eq.~(\ref{A.9}) is elementary:
\begin{equation}
\label{A.10}
 \int_{\eta^*}^{\eta_0} \; \Delta H(\eta') \;  d\eta'  \;  = \; \int_{\eta^*}^{\eta_0} \;  \frac{1}{2} \; \frac{d }{d \eta'} e^2(\eta')  \;  d\eta'  \;  = \; 
\int_{t^*}^{t_0} \;  \frac{1}{2} \; \frac{d }{d t'} e^2(t')  \; dt' \; = \;    \frac{1}{2} \;  e^2(t_0) \; - \;   \frac{1}{2} \;  e^2(t^*) \; = \;    - \;   \frac{1}{2} \;  e^2(t^*) \;,
\end{equation}
since $e^2(t_0) = 0$.  On the other hand, using Eq.~(\ref{A.3}) we have:
\begin{equation}
\label{A.11}
\frac{1}{7}  \int_{\eta_i}^{\eta^*} \Delta H(\eta') \left[ 6 e^{-\tau(\eta')} \,  +  \, e^{- \frac{3}{10} \tau(\eta')}  \right]  \;  d\eta'   \; \simeq \;
- \,  \frac{3}{4}  \int_{t_i}^{t^*}  \; e^2(t') \, H(t') \,   \left[  \frac{6}{7} \,  e^{-\tau(t')} \,  + \,  \frac{1}{7} \, e^{- \frac{3}{10} \tau(t')}  \right]   \;  d t' \; .
\end{equation}
After using  Eq.~(\ref{A.5}) we obtain:
\begin{equation}
\label{A.12}
\theta_a \;  \simeq \;  - \; \frac{1}{2}  \;  e^2_{\rm dec}  \; f(z^*)    \; ,
\end{equation}
where:
\begin{equation}
\label{A.13}
f(z^*) \;  = \;     (\frac{1 \, + \, z^*}{1 \, + z_d} )^{\frac{3}{2}}   \; + \; 
\frac{3}{2}  \int_{z^*}^{\infty}  \; (\frac{1 \, + \, z}{1 \, + z_d} )^{\frac{3}{2}}   \,
  \frac{1 }{1 + z}   \,   \left[  \frac{6}{7} \,  e^{-\tau(z)} \,  + \,  \frac{1}{7} \, e^{- \frac{3}{10} \tau(z)}  \right]   \;  d z \; .
\end{equation}
The integral in Eq.~(\ref{A.13}) can be evaluated numerically. In fact, we find that $f(z^*)$ is almost independent on $z^*$:
\begin{equation}
\label{A.14}
f(z^*) \;  \simeq \;   0.944  \;  \; , \; \; 200 \; \le \; z^* \; \le 900 \; .
\end{equation}
Thus, our final result is:
\begin{equation}
\label{A.15}
\theta_a \;  \simeq \;  - \; \frac{1}{2}   \times 0.944  \;  e^2_{\rm dec}  \; .
\end{equation}

\section{Multipole expansion of the large scale temperature anisotropies}
\label{Appendix B}
In this appendix we would like to discuss the multipole expansion of the temperature fluctuation correlation functions.
According to the results in sec.~\ref{S3}  we have (omitting the superscript A):
\begin{equation}
\label{B.1}
\Theta^T(\vec{k}, \vec{n},  \hat{p})  \; \simeq \;   \theta_a \,   (\cos^2 \theta_{\vec{n} \vec{ p}}  - \frac{1}{3})  \;  e^{- \, i \, k \,   \cos \theta_{\vec{k} \vec{ p}} \,  \eta_0} \;  \; ,
\end{equation}
\begin{equation}
\label{B.2}
\Theta^E(\vec{k},  \vec{n},  \hat{p})  \; \simeq \;   
 \theta_p \, (1 - \cos^2 \theta_{\vec{n} \vec{ p}}  ) \;  e^{- \, i \, k \,   \cos \theta_{\vec{k} \vec{ p}} \,  \eta_0}  \; \; ,
\end{equation}
corresponding to the temperature and polarization contrast functions, respectively. For definiteness, let us discuss firstly the temperature-temperature
correlations. As is well known~\citep{Dodelson:2003} we need to evaluate:
\begin{equation}
\label{B.3}
< \Theta^T(\vec{x}, \vec{n},  \hat{p}) \, \Theta^T(\vec{x}, \vec{n},  \hat{p}') >  \; = \;  \int \, \frac{d^3k}{(2 \pi)^3} \; \;
\Theta^T(\vec{k}, \vec{n},  \hat{p}) \, [\Theta^T(\vec{k}, \vec{n},  \hat{p}')]^* \;  \; .
\end{equation}
After expanding $\Theta^T(\vec{x}, \vec{n},  \hat{p})$ in spherical harmonics, one obtains:
\begin{equation}
\label{B.4}
< \Theta^T(\vec{x}, \vec{n},  \hat{p}) \, \Theta^T(\vec{x}, \vec{n},  \hat{p}') >  \; =  \;    \sum_{\ell}  \; C_{\ell}  \; \;  ,  \; \; 
C_{\ell} \; = \; \frac{1}{2 \ell+1} \sum_{m=-\ell}^{+\ell} \,  C_{\ell m}  \; \; ,
\end{equation}
with:
\begin{equation}
\label{B.5}
C_{\ell m}  \;  =  \;  \int \, \frac{d^3k}{(2 \pi)^3} \; \; \int \, d \Omega_{\hat{p}} \;  d \Omega_{\hat{p}'}  \;
Y^*_{\ell m}( \hat{p} )  \, \Theta^T(\vec{k}, \vec{n},  \hat{p}) \; Y_{\ell m}( \hat{p}') \,  [\Theta^T(\vec{k}, \vec{n},  \hat{p'})]^* 
\; \; .
\end{equation}
Now we use the well known identities~\citep{Abramowitz:1970}:
\begin{equation}
\label{B.6}
P_{\ell}(\hat{x} \cdot \hat{x}')  \;  =  \;   \frac{4 \, \pi}{ 2 \ell + 1}  \; 
\sum_{m = - \ell}^{m = + \ell} \; Y^*_{\ell m}( \hat{x} )  \; Y^*_{\ell m}( \hat{x}') \; \; , 
\end{equation}
and
\begin{equation}
\label{B.7}
 e^{- \, i \, \vec{k} \, \cdot  \, \vec{ x} } \; = \;  \sum_{\ell = 0}^{+ \infty} \; i^{\ell} \, ( 2 \ell + 1) \, j_{\ell}(k x) \, P_{\ell}(\hat{k} \cdot \hat{x})  \; = \; 
 4 \, \pi  \;   \sum_{\ell m} \;  i^{\ell} \;    j_{\ell}(k x) \; Y_{\ell m}( \hat{k} )  \; Y^*_{\ell m}( \hat{x})  \; \; ,
\end{equation}
to get:
\begin{equation}
\label{B.8}
C_{\ell m}  \;  =  \; \frac{8}{9 \pi} \, \theta_a^2  \, \sum_{\ell_1 m_1}  \int_0^{\infty}  \, dk \, k^2 \,  j^2_{\ell_1}(k \eta_0) 
\; \int \, d \Omega_{\hat{p}} \;  Y^*_{\ell m}( \hat{p} )  Y^*_{\ell_1 m}( \hat{p} )  P_{2}(\hat{p} \cdot \hat{n}) \;  
 \int \, d \Omega_{\hat{p}'}  \; Y_{\ell m}( \hat{p}' )  Y_{\ell_1 m}( \hat{p}' )  P_{2}(\hat{p}' \cdot \hat{n})   \; .
\end{equation}
Using again  Eq.~(\ref{B.6}) we rewrite   Eq.~(\ref{B.8}) as:
\begin{equation}
\label{B.9}
C_{\ell m}  \;  =  \; \frac{8}{9 \pi} \, (\frac{ 4 \pi}{5})^2 \; \theta_a^2  \, \sum_{\ell_1 m_1}  \sum_{ m_2 = - 2}^{+ 2}   \int_0^{\infty}  \, dk \, k^2 \,  j^2_{\ell_1}(k \eta_0) 
 \; Y_{2  m_2}( \hat{n} ) \;  Y^*_{2 m_2}( \hat{n} ) \;  
\;  \left | \;  \int \, d \Omega_{\hat{p}} \;  Y_{\ell m}( \hat{p} )  Y_{\ell_1 m_1}( \hat{p} )  Y_{2 m_2}(\hat{p}) \;  \right |^2  \;  
 \; .
\end{equation}
The angular integral can be expressed in terms of the Wigner 3j symbols~\citep{Messiah:1961}:
\begin{equation}
\label{B.10}
 \int \, d \Omega_{\hat{p}} \;  Y_{\ell_1 m_1}( \hat{p} )  Y_{\ell_2 m_2}( \hat{p} )  Y_{\ell_3 m_3}(\hat{p}) \;  =  \; 
 \sqrt{ \frac{ (2 \ell_1 +1) \, (2 \ell_2 +1) \,  (2 \ell_3 +1)}{4 \, \pi} } \;  
   \pmatrix{\ell_1 \; \ell_2 \; \ell_3 \cr 0 \; \;  0 \; \;  0 \cr}  \;    \pmatrix{\ell_1 \; \; \ell_2 \; \; \ell_3 \cr m_1 \; m_2 \; m_3 \cr}  \; .
\end{equation}
In our case, using the well-known properties of the 3j symbols, we have the constraints:
\begin{equation}
\label{B.11}
\ell_1 \; = \; \ell \; , \; \ell \; \pm \; 2 \; .
\end{equation}
Actually, we are interested in the limit of large $\ell$. To this end, we use the estimate for the asymptotic limit of the average
3j symbols ~\citep{Borodin:1978}:
\begin{equation}
\label{B.12}
 \left < \pmatrix{\ell_1 \; \;  \;  2  \; \; \; \ell  \cr m_1 \; m_2 \; m  \cr}^2 \right >  \;  \simeq \; \frac{1}{2 \, \pi \, \ell^2} \; \;  \; ,  \; \; \ell_1 \; \simeq \; \ell \; , \; m_1 \; \simeq \; m \; ,
\end{equation}
to get:
\begin{equation}
\label{B.13}
C_{\ell m}  \;  \simeq  \; \frac{8}{9 \pi} \; \theta_a^2  \frac{5}{4 \, \pi^3} \; \frac{1}{\ell^2} \;   \int_0^{\infty}  \, dk \, k^2 \,  j^2_{\ell}(k \eta_0) 
 \;   \; (\frac{ 4 \pi}{5})^2   \sum_{ m_2 = - 2}^{+ 2}  \, Y_{2  m_2}( \hat{n} ) \;  Y^*_{2 m_2}( \hat{n} ) \;   \;    \; .
\end{equation}
Since 
\begin{equation}
\label{B.14}
  \frac{ 4 \pi}{5}    \sum_{ m_2 = - 2}^{+ 2}  \, Y_{2  m_2}( \hat{n} ) \;  Y^*_{2 m_2}( \hat{n} ) \;   = \; P_2(1) \;  =  \;  1    \; ,
\end{equation}
we obtain:
\begin{equation}
\label{B.15}
C_{\ell} \; = \; \frac{1}{2 \ell+1} \sum_{m=-\ell}^{+\ell} \,  C_{\ell m}  \;  \simeq \;
 \frac{8}{9 \, \pi^3} \; \theta_a^2  \; \frac{1}{\ell^2} \;   \int_0^{\infty}  \, dk \, k^2 \,  j^2_{\ell}(k \eta_0)  \; .
\end{equation}
To evaluate the integral over $k$, we note that:
\begin{equation}
\label{B.16}
j_{\ell}(x)  \; = \; \sqrt{ \frac{\pi}{2 \, x}} \; J_{\ell+\frac{1}{2}}(x) \; . 
\end{equation}
So that we are left with the following integrals:
\begin{equation} 
\label{B.17}
I_{\ell}  \; \equiv \;    \int_0^{\infty}  \, dk \, k^2 \,  j^2_{\ell}(k \eta_0)  \; 
= \;   \frac{\pi}{2} \;  \int_0^{\infty}  \, dk \, \frac{k}{\eta_0}  \,  J^2_{\ell + \frac{1}{2}}(k \eta_0) \; .
\end{equation}
It is easy to see that the integrals in Eq.~(\ref{B.17}) are divergent in the ultraviolet region $k \rightarrow \infty$. This divergence
is an artifact of our approximations. To overcome this problem we must cut-off the spectrum for high wavenumbers. 
To our purpose it is enough to assume a power law cut-off function $k^{-\alpha} \; , \; 0 < \alpha < 1$.
 Thus we obtain:
\begin{equation} 
\label{B.18}
I_{\ell}  \; = \;    \frac{\pi}{2 \,  \eta_0^{3 - \alpha}} \;  \int_0^{\infty}  \, dx \, x^{1 - \alpha}  \,  J^2_{\ell + \frac{1}{2}}(x) \; .
\end{equation}
Using~\citep{Gradshteyn:1983}:
\begin{equation} 
\label{B.19}
\int_0^{\infty}  \, dt \, t^{- \lambda}  \,  J_{\nu}(t)  \,  J_{\mu}(t) \; = \; 
\frac{ \Gamma(\lambda) \; \Gamma(\frac{\mu  + \nu - \lambda + 1}{2})}{ 2^{\lambda} \,  \Gamma( \frac{\mu -\nu + \lambda + 1}{2}) \,
 \Gamma(\frac{\mu +  \nu + \lambda + 1}{2}) \,  \Gamma(\frac{- \mu  + \nu + \lambda + 1}{2})} \; ,
\end{equation}
we obtain:
\begin{equation} 
\label{B.20}
I_{\ell}  \; = \;    \frac{\pi}{2 \,  \eta_0^{3 - \alpha}} \;    \frac{1}{2^{\alpha - 1}}       \; \frac{ \Gamma(\alpha - 1)}{[\Gamma(\frac{\alpha}{2})]^2} \; 
\frac{ \Gamma(\ell + \frac{3}{2}  - \frac{\alpha}{2})}{ \Gamma(\ell + \frac{1}{2}   + \frac{\alpha}{2}) } \; .
\end{equation}
For large $\ell$ we use the estimate~\citep{Gradshteyn:1983}:
\begin{equation} 
\label{B.21}
\lim_{ |z| \rightarrow \infty}  \frac{\Gamma(z + a)}{\Gamma(z)} \; = \;  e^{- a \ln |z|} \; = \; |z|^{-a} \; ,
\end{equation}
to obtain:
\begin{equation} 
\label{B.23}
I_{\ell}  \; \sim \;    \frac{\pi}{2^{\alpha}  \,  \eta_0^{3 - \alpha}} \;      \; \frac{ \Gamma(\alpha - 1)}{[\Gamma(\frac{\alpha}{2})]^2} \; 
\ell^{-1 + \alpha}  \; .
\end{equation}
Inserting Eq.~(\ref{B.23})   Eq.~(\ref{B.15}) we get:
\begin{equation}
\label{B.24}
\frac{\ell (\ell + 1)}{2 \, \pi} \; C_{\ell} \; \sim  \;  \frac{8}{9 \, \pi^2} \; \theta_a^2  \; \frac{1}{ 2^{\alpha}  \,  \eta_0^{3 - \alpha }} \;
\frac{ \Gamma(\alpha - 1)}{[\Gamma(\frac{\alpha}{2})]^2} \; \ell^{-1 + \alpha}  \;  \; , \; \;  0 \; < \alpha \; < \; 1 \; .
\end{equation}
This last equation shows, indeed, that the anisotropy of the metric contributes mainly at large scales affecting only
the low-lying multipoles, at least for the temperature-temperature anisotropy correlations. \\
For the polarization correlations, we rewrite Eq.~(\ref{B.2})  as:
\begin{equation}
\label{B.25}
\Theta^E(\vec{k},  \vec{n},  \hat{p})  \; \simeq \;   
 \frac{2}{3} \, \theta_p  \;  e^{- \, i \, k \,   \cos \theta_{\vec{k} \vec{ p}} \,  \eta_0}  \; 
 - \,  \frac{2}{3} \,  \theta_p \, P_2( \cos \theta_{\vec{n} \vec{ p}} ) \;  e^{- \, i \, k \,   \cos \theta_{\vec{k} \vec{ p}} \,  \eta_0}  \;  .
\end{equation}
Therefore, we have two contributions to the EE correlations. The second term on the right hand of  Eq.~(\ref{B.25}) is
analogous to the temperature contrast function Eq.~(\ref{B.1}), while the first term would contribute to the coefficient 
$C_{\ell}$ with:
\begin{equation}
\label{B.26}
C_{\ell} \;   \simeq \;
 \frac{8}{9 \, \pi} \; \theta_p^2   \;   \int_0^{\infty}  \, dk \, k^2 \,  j^2_{\ell}(k \eta_0)  \; .
\end{equation}
Using Eqs.~(\ref{B.19}) - (\ref{B.21}) we find:
\begin{equation}
\label{B.27}
\frac{\ell (\ell + 1)}{2 \, \pi} \; C_{\ell} \; \sim  \;  \frac{8}{9 } \; \theta_p^2  \; \frac{1}{ 2^{\alpha}  \,  \eta_0^{3 - \alpha }} \;
\frac{ \Gamma(\alpha - 1)}{[\Gamma(\frac{\alpha}{2})]^2} \; \ell^{1 + \alpha}  \;  \; , \; \;  0 \; < \alpha \; < \; 1 \; .
\end{equation}
Eq.~(\ref{B.27}) would imply that the EE correlation functions due to the  anisotropy of the metric are   sizable not only for
 the low-lying multipoles, but also for higher multipoles. However, from Eq.~(\ref{3.18})   we see  that the polarization 
 of the cosmic microwave background (without reionization) at the present time is essentially 
that produced around the time of recombination, since much later the free electron density is negligible,  
while much earlier the optical depth is very large. Then, the present polarization  is the result of Thomson scattering
 around the time of decoupling of matter and radiation, which occurs after the free electron density starts to drop significantly~\citep{Peebles:1993}.  
Moreover,  to obtain Eq.~(\ref{3.19}) we assumed that $k \, \Delta \eta_d \; \ll \; 1$, where $\Delta  \eta_d$ is the conformal time 
duration of the  decoupling process (the thickness of the last scattering surface). In fact, for $k \, \Delta \eta_d \; \gg \; 1$
the oscillations in the integrand produce a cancellation of the temperature anisotropy polarization. In other words,
the finite thickness of the last scattering surface damps the final temperature polarization on these scales.
Thus, for wavelengths comparable  or smaller than the width of the last scattering surface, the polarization should fall off very rapidly. 
Indeed, the polarization signal should be confined up to multipoles   such that $\frac{1}{\ell}  \sim \frac{ \Delta z_d}{ z_d}  \sim 10^{-1}$.  
 Actually,  more precise statements can be only obtained by solving numerically the radiative transfer equation 
 for the cosmic microwave background including polarization in anisotropic universes. Remarkably, quite recently~\citet{Pontzen:2007}   have 
derived the  radiative transfer equation  in the nearly Friedmann-Robertson-Walker limit of homogeneous, 
but anisotropic, universes classified via their Bianchi type. In fact,  these authors argued that  the  polarization signal is mostly 
confined to multipoles $\ell \lesssim 10$.

\label{lastpage}


\begin{thebibliography}{}
%
\bibitem[\protect\citeauthoryear{Abramowitz \& Stegun}%
{1970}]{Abramowitz:1970} Abramowitz  M.,  Stegun  I. A.,  1970,  Handbook of Mathematical Functions, Dover Publications, New York
%
\bibitem[\protect\citeauthoryear{Ade et al.}%
{2013a}]{Ade:2013a} Ade P.A.R.,  et al., Planck Collaboration, 2013a,  arXiv:1303.5062 [astro-ph.CO]
%
\bibitem[\protect\citeauthoryear{Ade et al.}%
{2013b}]{Ade:2013b} Ade P.A.R.,  et al., Planck Collaboration, 2013b,  arXiv:1303.5072 [astro-ph.CO]
%
\bibitem[\protect\citeauthoryear{Ade et al.}%
{2013c}]{Ade:2013c} Ade P.A.R.,  et al., Planck Collaboration, 2013c,  arXiv:1303.5075 [astro-ph.CO]
%
\bibitem[\protect\citeauthoryear{Ade et al.}%
{2013d}]{Ade:2013d} Ade P.A.R.,  et al., Planck Collaboration, 2013d,  arXiv:1303.5076 [astro-ph.CO]
%
\bibitem[\protect\citeauthoryear{Ade et al.}%
{2013e}]{Ade:2013e} Ade P.A.R.,  et al., Planck Collaboration, 2013e,  arXiv:1303.5083 [astro-ph.CO]
%
%
\bibitem[\protect\citeauthoryear{Basko \& Polnarev}%
{1980}]{Basko:1980} Basko M.~M., Polnarev A.~G.,  1980, MNRAS,  191, 207
%
\bibitem[\protect\citeauthoryear{Bennett et al.}%
{2013}]{Bennett:2013} Bennett C.L.,  et al., 2013,  ApJS, 208, 20
%
\bibitem[\protect\citeauthoryear{Borodin, Kroshilin \& Tolmachev}%
{1978}]{Borodin:1978} Borodin  K. S.,  Kroshilin  A. E., Tolmachev V. V., 1978,   TMP,  34, 69
%
%
\bibitem[\protect\citeauthoryear{Campanelli, Cea \& Tedesco}%
{2006}]{Campanelli:2006} Campanelli L.,  Cea P., Tedesco L., 2006,   PRL,  97, 131302 ; (E) 209903
%
\bibitem[\protect\citeauthoryear{Campanelli, Cea \& Tedesco}%
{2007}]{Campanelli:2007} Campanelli L.,  Cea P., Tedesco L., 2007,   PRD,  76, 063007
%
\bibitem[\protect\citeauthoryear{Cea}%
{2010}]{Cea:2010}   Cea P.,  2010, MNRAS,  406, 586
%
%
\bibitem[\protect\citeauthoryear{Chandrasekhar}%
{1960}]{Chandrasekhar:1960} Chandrasekhar S., 1960,  Radiative Transfer, Dover Publications, New York
%
\bibitem[\protect\citeauthoryear{Copi et al.}%
{2006}]{Copi:2006} Copi C. J., Huterer D.,  Schwarz D. J.,  Starkman G. D.,  2006,
 MNRAS, 367, 79
%
%
\bibitem[\protect\citeauthoryear{Cruz et al.}%
{2005}]{Cruz:2005} Cruz M., Martinez-Gonzalez E., Vielva P.,  Cayon L.,
 2005, MNRAS, 356, 29
%
%
\bibitem[\protect\citeauthoryear{Dodelson}%
{2003}]{Dodelson:2003}   Dodelson S., 2003,  Modern Cosmology,  Academic Press, San Diego, California
%
%
\bibitem[\protect\citeauthoryear{Eriksen et al.}%
{2004}]{Eriksen:2004}  Eriksen, H. K.,  Hansen, F. K.,  Banday, A. J.,  Gorski, K. M., 
 Lilje, P. B.,  2004,  ApJ,  605, 14
%
%
\bibitem[\protect\citeauthoryear{Fixsen}%
{2009}]{Fixsen:2009} Fixsen D. J.,  2009,  ApJ, 707, 916
%
%
\bibitem[\protect\citeauthoryear{Gradshteyn \& Ryzhik}%
{1983}]{Gradshteyn:1983}  Gradshteyn I. S., Ryzhik I. M.,  1983,  Table of Integrals, Series, and
Products,   Academic Press, London
%
%
\bibitem[\protect\citeauthoryear{Hansen,  Banday \& Gorski }%
{2004}]{Hansen:2004}  Hansen F. K., Banday A. J.,  Gorski K. M.,  2004,  MNRAS,  354,  641
%
%
\bibitem[\protect\citeauthoryear{Hinshaw et al.}%
{2007}]{Hinshaw:2007} Hinshaw G. F., et al.,  2007,  ApJS, 170, 288
%
\bibitem[\protect\citeauthoryear{Hinshaw et al.}%
{2013}]{Hinshaw:2013} Hinshaw G.F.,  et al., 2013,  ApJS, 208, 19
%
%
\bibitem[\protect\citeauthoryear{Jones \& Wyse }
{1985}]{Jones:1985} Jones B.~J.~T., and Wyse R.~F.~G., 1985, A\&A, 149, 144
%
%
\bibitem[\protect\citeauthoryear{Land \& Magueijo}%
{2005}]{Land:2005} Land K., Magueijo J., 2005, PRL, 95, 071301
%
%
\bibitem[\protect\citeauthoryear{de Oliveira-Costa et al.}%
{2004}]{Oliveira:2004} de Oliveira-Costa A., Tegmark M., Zaldarriaga M., Hamilton A.,
 2004, PRD, 69, 063516
%
%
\bibitem[\protect\citeauthoryear{de Oliveira-Costa \& Tegmark}%
{2006}]{Oliveira:2006} de Oliveira-Costa A.,  Tegmark M., 2006,  PRD,  74, 023005
%
%
\bibitem[\protect\citeauthoryear{Messiah}%
{1961}]{Messiah:1961} Messiah  A.,  1961,  Quantum Mechanics,  North-Holland, Amsterdam
%
%
\bibitem[\protect\citeauthoryear{Mukhanov}%
{2005}]{Mukhanov:2005}   Mukhanov  V., 2005, Physical Foundation of Cosmology,  Cambridge University Press, New York
%
\bibitem[\protect\citeauthoryear{Negroponte \& Silk}%
{1980}]{Negroponte:1980} Negroponte J.,  Silk J., 1980,  PRL,   44, 1433
%
%
\bibitem[\protect\citeauthoryear{Park, Park \& Gott }
{2007}]{Park:2007} Park C.~G., Park C., and Gott J.~R.~I., 2007, ApJ, 660, 959
%
\bibitem[\protect\citeauthoryear{Peebles}%
{1993}]{Peebles:1993} Peebles P. J. E., 1993, Principles of Physical Cosmology, Princeton University Press, Princeton
%
%
\bibitem[\protect\citeauthoryear{Pontzen \&  Challinor}%
{2007}]{Pontzen:2007}  Pontzen A.,  Challinor A., 2007,  MNRAS,  380, 1387
%
\bibitem[\protect\citeauthoryear{Ralston \& Jain}%
{2004}]{Ralston:2004} Ralston J. P., Jain P., 2004, Int. J. Mod. Phys. D, 13, 1857
%
%
\bibitem[\protect\citeauthoryear{Rees}%
{1968}]{Rees:1968} Rees M.~J., 1968,   ApJ, 153, L1
%
%
\bibitem[\protect\citeauthoryear{Vilenkin \& Shellard}%
{1994}]{Vilenkin:1994} Vilenkin A., Shellard E. P. S., 1994, 
Cosmic Strings and Other Topological Defects, Cambridge University Press, Cambridge
%
%
\end{thebibliography}
\end{document}